\documentclass[preprint,3p,12pt,sort&compress]{elsarticle}
\usepackage{mathrsfs}
\usepackage{amsmath}
\usepackage{stmaryrd}
\usepackage{bbding}
\usepackage{dcolumn}
\usepackage{graphicx}
\usepackage{amsfonts}
\usepackage{amssymb}
\usepackage{psfrag}
\usepackage{wrapfig}
\usepackage{subfigure}
\usepackage{makeidx}
\usepackage{bm}
\usepackage{epsf}
\usepackage{epsfig}
\usepackage{setspace}
\usepackage{graphicx}
\usepackage{epstopdf}
\usepackage{psfrag}
\usepackage{subfigure}
\usepackage{color}
\usepackage{enumerate}
\usepackage{verbatim} 
\usepackage{booktabs}

\begin{document}

\title{Using neural networks to accelerate the solution of the Boltzmann equation}

\author[KIT]{Tianbai Xiao\corref{cor}}
\ead{tianbaixiao@gmail.com}

\author[KIT]{Martin Frank}
\ead{martin.frank@kit.edu}

\address[KIT]{Karlsruhe Institute of Technology, Karlsruhe, Germany}
\cortext[cor]{Corresponding author}

\begin{abstract}

    One of the biggest challenges for simulating the Boltzmann equation is the evaluation of fivefold collision integral. Given the recent successes of deep learning and the availability of efficient tools, it is an obvious idea to try to substitute the evaluation of the collision operator by the evaluation of a neural network. However, it is unlcear whether this preserves key properties of the Boltzmann equation, such as conservation, invariances, the H-theorem, and fluid-dynamic limits.
    
    In this paper, we present an approach that guarantees the conservation properties and the correct fluid dynamic limit at leading order. The concept originates from a recently developed scientific machine learning strategy which has been named ``universal differential equations''.
    It proposes a hybridization that fuses the deep physical insights from classical Boltzmann modeling and the desirable computational efficiency from neural network surrogates.
    The construction of the method and the training strategy are demonstrated in detail. We conduct an asymptotic analysis and illustrate its multi-scale applicability. The numerical algorithm for solving the neural network-enhanced Boltzmann equation is presented as well.
    Several numerical test cases are investigated.
    The results of numerical experiments show that the time-series modeling strategy enjoys the training efficiency on this supervised learning task.
    
\end{abstract}


\maketitle

\section{Introduction}

Modern data-driven techniques widen the possibility of solving the problems that seemed beset with difficulties in the past, e.g., computer vision \cite{nixon2019feature} and natural language processing \cite{chowdhary2020natural}.
The same momentum is building in computational sciences, leading to the so-called scientific machine learning \cite{mjolsness2001machine}.
However, given the high expense of conducting experiments and numerical simulations, e.g.\ in fluid dynamical and astronautical research, it is challenging to establish an all-round data base. While the generalization performance of neural networks based on small training sets is questionable, quantitative interpretability lies at the core of scientific modeling and simulation, which more or less collides with the blackbox nature of multi-layer neural networks.

Several approaches that try to combine the advantages of differential equations and machine learning have emerged recently. Physics-informed neural networks (PINNs) directly incorporate the structure of differential equations into cost function and train the solutions in terms of neural networks \cite{raissi2019physics}.
The sparse identification of nonlinear dynamical systems (SINDy) \cite{brunton2016discovering} employs sparse regression to select most probable equations from data.
These two methods provide efficient tools for identifying and solving ordinary and partial differential equations, e.g.\ the Navier-Stokes equations, and require relatively a small amount of data.

As we look into multi-scale fluid mechanics with upscaling effects from the atomistic level, more complex dynamical systems could emerge. A typical example is the Boltzmann equation, which describes the evolution of the one-particle probability density function $f(t, \mathbf{x}, \mathbf{u})$, which describes the probability of finding a particle with a certain location $\mathbf{x}$ and speed $\mathbf u$.
In the absence of external force field, the Boltzmann equation reads as follows,
\begin{equation}
    \frac{\partial f}{\partial t}+\mathbf{u} \cdot \nabla_{\mathbf{x}} f = Q(f)=\int_{\mathcal R^3} \int_{\mathcal S^2} \mathcal B(\cos \beta, g) \left[ f(\mathbf u')f(\mathbf u_*')-f(\mathbf u)f(\mathbf u_*)\right] d\mathbf \Omega d\mathbf u_*,
    \label{eqn:boltzmann}
\end{equation}
where $\{\mathbf{u},\mathbf{u_*}\}$ are the pre-collision velocities of two colliding particles, and $\{\mathbf{u}',\mathbf{u_*}'\}$ are the corresponding post-collision velocities.
The collision kernel $\mathcal B(\cos \beta,g)$ measures the strength of collisions in different directions, where $g=|\mathbf g|=|\mathbf{u}-\mathbf{u_*}|$ is the magnitude of relative pre-collision velocity, $\mathbf{\Omega}$ is the unit vector along the relative post-collision velocity $\mathbf{u}'-\mathbf{u_*}'$,
and the deflection angle $\beta$ satisfies the relation $\cos \beta = \mathbf \Omega \cdot \mathbf g/g$.

The Boltzmann equation serves as the basis of many high-level theories, e.g.\ non-equilibrium thermodynamics and extended hydrodynamics.
As shown, the Boltzmann equation is an integro-differential equation, with its right-hand side being a fivefold integral over phase space.
This convolution-type collision operator brings tremendous difficulty to the application of PINN since a direct differentiable structure is absent.
On the other hand, despite the development of classical numerical solvers, the computational cost of solving the Boltzmann collision integral can be prohibitive.
Consider the fast spectral method \cite{mouhot2006fast}, an efficient Boltzmann solution algorithm which employs Fast Fourier Transformation to compute convolutions within spectral space.
The computational cost of it is $O(M^2 N_x^D N_u^D \log N_u)$, where $N_x$, $N_u$ and $M$ are the numbers of grids in physical, velocity and angular space with dimension $D$ \cite{xiao2019unified}.
Obviously it's still unrealistic to perform a direct numerical simulation for a real-world application in aerospace industry.
Moreover, due to the high-dimensional nature of intermolecular interactions, it is sometimes cumbersome to adopt the Boltzmann solver if we are interested in one-dimensional distribution of solutions only, e.g. the profiles of macroscopic variables inside a shock tube.

The paper is organized as follows.
In Sec.\ref{sec:kinetic} we introduce some fundamental concepts in the kinetic theory of gases.
Sec.\ref{sec:nde} presents the main idea of this work, and
Sec.\ref{sec:scheme} details the numerical solution algorithm.
Sec.\ref{sec:experiment} contains the numerical experiments for both spatially homogeneous and inhomogeneous cases to validate the current method.
The last section is the conclusion.

\section{\label{sec:kinetic}Kinetic theory of gases}

Kinetic theory provides a one-to-one correspondence with its macroscopic limit system.
Taking moments through particle velocity space, we get the macroscopic mass, momentum and energy density,
\begin{equation}
    \mathbf{W} (t, \mathbf x) =\left(
    \begin{matrix}
    \rho \\
    \rho \mathbf U \\
    \rho E
    \end{matrix}
    \right)=\int f\psi d\mathbf u,
    \label{eqn:macro define}
\end{equation}
where $\psi=\left(1,\mathbf u,\frac{1}{2} \mathbf u^2 \right)^T$ is a vector of collision invariants.
The collision operator satisfies the compatibility condition for conservative variables, i.e.,
\begin{equation}
    \int Q(f)\psi d\mathbf u=0.
    \label{eqn:compatibility condition}
\end{equation}
Substituting the $H$ function,
\begin{equation*}
	H(t, \mathbf x) = - \int f \ln f d \mathbf{u},
\end{equation*}
into the Boltzmann equation we have
\begin{equation}
\frac{\partial H}{\partial t} =-\int(1+\ln f) \frac{\partial f}{\partial t} d \mathbf{u} =-\iiint(1+\ln f)\left(f^{\prime} f_{*}^{\prime}-f f_{*}\right) \mathcal B d \Omega d \mathbf{u} d \mathbf{u}_{*}.
\end{equation}
From the H-theorem \cite{cercignani1988boltzmann} we know that entropy is locally maximal when $f$ is a Maxwellian
\begin{equation}
\mathcal M(t, \mathbf x, \mathbf u, \mathbf z)=\rho \left( \frac{\lambda}{\pi} \right)^{\frac{3}{2}} e^{-\lambda(\mathbf u-\mathbf U)^2},
\label{eqn:maxwell distribution}
\end{equation}
where $\lambda=m/(2kT)$, $m$ is molecule mass and $k$ is the Boltzmann constant.

Since intermolecular collisions drive the system towards Maxwellian, simplified relaxation models, e.g. the Bhatnagar-Gross-Krook (BGK) \cite{bhatnagar1954model} and Shakhov \cite{shakhov1968generalization}, have been constructed.
It writes 
\begin{equation}
    \frac{\partial f}{\partial t}+\mathbf{u}\cdot\nabla_{\mathbf x} f = Q(f) = \nu(f^+-f), \\
    \label{eqn:bgk equation}
\end{equation}
where $\nu$ is collision frequency.
For the BGK model, the equilibrium state is Maxwellian $f^+=\mathcal M$, while in the Shakhov model it takes the form
\begin{equation}
    f^{+}=\mathcal{M}\left[1+(1-\mathrm{Pr}) (\mathbf{u-U}) \cdot \mathbf q\left(\frac{(\mathbf{u-U})^{2}}{R T}-5\right) /(5 p R T)\right],
    \label{eqn:shakhov}
\end{equation}
where Pr is the Prandtl number, $\mathbf q$ is heat flux, $p$ is pressure and $R$ is gas constant.
The relaxation models avoid the complicated fivefold Boltzmann integral.
They still possesses some key properties of the original Boltzmann equation, e.g. the H-theorem, but fail to provide exactly equivalent Boltzmann solutions as as the distribution function deviates far from the Maxwellian.

\section{\label{sec:nde}Neural Network-Enhanced Boltzmann equation}

\subsection{Idea}
One intuitive strategy to make use of neural networks is to replace the right-hand-side of the Boltzmann equation with a neural network directly. Borrowing from the neural ordinary differential equation (ODE) \cite{chen2018neural}, we call this neural Boltzmann equation (NBE):
\begin{equation}
    f_t=\mathrm{NN}_\theta(f, t),
\end{equation}
where $\theta$ denotes the collection of all the parameters inside neural network ($\rm NN$).

Neural ODEs are a family of deep neural network models.
This idea originates from the structure of some state-of-the-art neural networks, e.g. residual neural network (ResNet), which updates the hidden layer with the strategy
\begin{equation}
    \mathbf h^{n+1}=\mathbf h^n+\mathcal F(\mathbf h^n,\theta^n).
    \label{eqn:resnet}
\end{equation}
Such an iterative stepping is equivalent to the forward Euler method of a differential equation.
Therefore, in the limiting case with infinitely small time step, the discrete iteration can be concluded by an ordinary differential equation in terms of neural network, i.e. the so-called neural ODE,
\begin{equation}
    \mathbf h_t=\mathcal F_\theta(\mathbf h,t).
    \label{eqn:neural ode}
\end{equation}
It forms an initial value problem (IVP) for the hidden layers, where modern ODE solvers can be used with monitoring of accuracy and efficiency.
Since the derivative of the hidden state is parameterized with continuous dynamics, the parameters of original discrete sequence layers in Eq.(\ref{eqn:resnet}) can be regarded as seamlessly coupled.
As a result, for a typical supervised learning task, the required number of parameters drops correspondingly \cite{chen2018neural}.
Modern ODE solvers can be employed to solve the IVP with monitoring of desirable accuracy and efficiency.
No intermediate quantities of forward pass need to be stored, leading to a constant memory cost as a function of depth.
Also, the continuous modeling make it much easier to perform interpolation and extrapolation beyond the training data.

In spite of the advantages, due to the blackbox nature of neural networks, this approach does not guarantee any property of the Boltzmann equation when the training isn't perfect. 
One rather general approach to enforce physical constraints has been presented under the name universal differential equations (UDE) \cite{rackauckas2020universal}, in which the model is a combination of mechanical and neural parts.
Continuing with the example above, we rewrite the Boltzmann equation into a universal Boltzmann equation (UBE) as
\begin{equation}
    f_t=Q(f, t, \mathrm{NN}_\theta(f,t)),
\end{equation}
where $Q$ is the particle collision term, and $\mathrm{NN}_\theta(f,t)$ denotes the neural network model that plays a portion necessary for the self-contained physical description but missed from the mechanical modeling.

The key idea to go beyond the mere approxiamtion of the right-hand side (and thus to obtain a neural differential equation) is to split the right-hand side into a mechanisitc part and a part to be approximated. In this paper, we employ the BGK equation as the mechanical part of UBE, leaving the difference to the full Boltzmann collision operator to be approximated by a neural network.
The BGK equation provides a lightweight model that balances physical insight and numerical efficiency.
It holds a similar structure as Boltzmann equation, while the computational cost is of $O(N^D)$, where $N$ is the number of discrete velocity grids and $D$ is dimension.
Therefore, it is significantly more efficient than solving the full Boltzmann integral.
The concrete UBE is designed as follows,
\begin{equation}
    \frac{\partial f}{\partial t}+\mathbf{u} \cdot \nabla_{\mathbf{x}} f =\nu (\mathcal M - f) + \mathrm{NN}_\theta(\mathcal M - f).
\end{equation}
where $\mathrm{NN}_\theta$ can be a concrete type of neural network,
with the input set as the difference between the Maxwellian and current particle distribution function.

The proposed UBE has the following benefits.
First, it automatically ensures the asymptotic limits.
Let us consider the Chapman-Enskog method for solving Boltzmann equation \cite{chapman1970mathematical}, where the distribution function is approximated with expansion series,
\begin{equation}
    f \simeq f^{(0)} + f^{(1)} + f^{(2)}+\cdots, \quad f^{(0)}=\mathcal M.
\end{equation}
Take the zeroth order truncation, and consider an illustrative multi-layer perceptron with 
\begin{equation}
    \mathrm{NN}_\theta(x)=\mathrm{layer}_n (\dots \mathrm{layer}_2(\mathcal A(\mathrm{layer}_1( x)))), \quad \mathrm{layer}(x)=w  x,
\end{equation}
where each layer is a matrix multiplication followed by activation function $\mathcal A$ that can adpot sigmoid, tanh, etc.
Given the zero input from $\mathcal M-f$, the contribution from collision term is absent.
Taking moments with respect to collision invariants,
\begin{equation}
    \int\left(\begin{array}{c}
    1 \\
    \mathbf u \\
    \frac{1}{2} \mathbf u^{2}
    \end{array}\right)\left( {\mathcal M}_{t}+\mathbf u \cdot \nabla_\mathbf x {\mathcal M} \right) d \mathbf u=0,
\end{equation}
we arrive at the corresponding Euler equations,
\begin{equation}
    \frac{\partial}{\partial t}\left(\begin{array}{c}
    \rho \\
    \rho \mathbf U \\
    \rho E
    \end{array}\right)+
    \nabla_{\mathbf x} \cdot \left(\begin{array}{c}
    \rho \mathbf U \\
    \rho \mathbf U \otimes \mathbf U \\
    \mathbf U (\rho E + p)
    \end{array}\right)=0.
\end{equation}
As is shown, the asymptotic property of UBE in the hydrodynamic limit is preserved independent of the training parameters $\theta$.

\begin{figure}[htb!]
	\centering
	\includegraphics[width=0.7\textwidth]{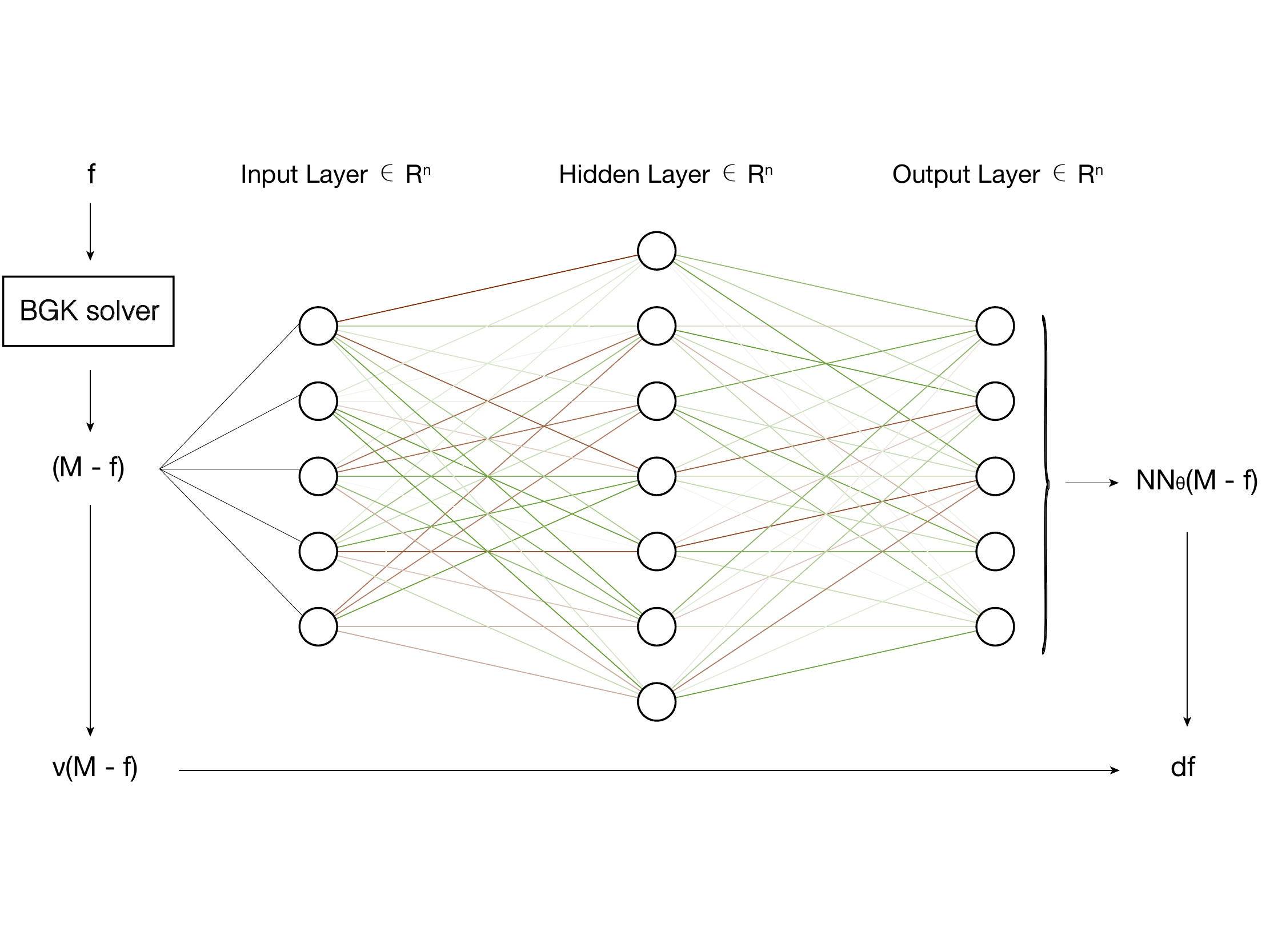}
	\caption{An illustrative flow chart for the collision term evaluation in the universal Boltzmann equation.}
	\label{pic:nn}
\end{figure}

Another advantage from current strategy is the training efficiency.
Since the BGK relaxation term provides a qualitatively mechanism to describe gas evolution, as analyzed in \cite{xiao2020velocity}, after the evolving time from initial strong non-equilibrium exceed a few collision time, the difference between Boltzmann integral and BGK model becomes minor.
Therefore, now the task left becomes to train a neural network that approximates solutions close to zero, which will significantly accelerate the convergence of $\theta$.
Fig.\ref{pic:nn} provides an illustration for collision term evaluation in the universal Boltzmann equation.
The detailed training strategy will be presented in the next subsection.

\subsection{Training strategy}

Training NBE and UBE with datasets consisting of exact or reference solutions is a typical supervised learning.
It amounts to an optimization problem which minimizes the difference between the current predictions and ground-truth solutions.
For example, a cost function can be defined based on the Euclidean distance along discrete grid points,
\begin{equation}
    C(\theta)=\sum_{i,j,n} ||f_\theta-f_{\mathrm{ref}} ||(t^n,\mathbf x_i,\mathbf u_j).
\end{equation}

Generally, the optimization algorithms can be classified into gradient-free and gradient-required methods.
Thanks to the rapid development of automatic differentiation (AD), the latter one becomes prevalent in machine learning community.
There are two modes of AD, i.e. the forward-mode and the reverse-mode, which differ from the direction of evaluating the chain rules.
Here we focus on the latter.
Consider a smooth function $y = \mathcal F(x)$, 
the reverse-mode AD computes the dual (conjugate-transpose) matrix of Jacobian $\mathcal J=\nabla \mathcal F$ at $x=x_0$ with the chain rule,
\begin{equation}
    \left(\mathcal{J}(\mathcal F)\left(x_{0}\right)\right)^{*}=\left(\mathcal{J}\left(G_{1}\right)\left(x_{0}\right)\right)^{*} \times \cdots \times\left(\mathcal{J}\left(G_{k}\right)\left(x_{k-1}\right)\right)^{*},
\end{equation}
with $x_{i}:=G_{i}\left(x_{i-1}\right) \text { for } i=1, \ldots, k-1$.

As the reverse-mode AD can naturally be expressed using pullbacks and differential one-forms from geometric perspective,
in this work we employ open-source package Zygote.jl \cite{innes2019differentiable}, which utilizes pullback functions to perform reverse-mode AD.
Different from the tracing methods used in Tensorflow \cite{abadi2016tensorflow} and PyTorch \cite{paszke2017automatic}, it employs the source-to-source mode via differentiable programming, i.e. generates derivative directly from pullback functions.
Such an approach enjoys the benefits of, e.g. low overhead, efficient support for control flow and user-defined data types and dynamism.

When the derivatives of cost function are evaluated by automatic differentiation, gradient-descent-type optimizers can be employed, e.g. stochastic gradient decent, ADAM \cite{kingma2014adam}, Nesterov \cite{nesterov2012efficiency},  Broyden–Fletcher–Goldfarb–Shanno (BFGS) \cite{nocedal2006numerical}, or its limited-memory version (L-BFGS).
In this paper, we adopt the scientific machine learning framework in DiffEqFlux.jl \cite{rackauckas2019diffeqflux} for training neural networks.

The training set is produced by the fast spectral method \cite{wu2013deterministic} with respect to different initial values of homogeneous Boltzmann equation
\begin{equation}
    f_t=\int_{\mathcal R^3} \int_{\mathcal S^2} \mathcal B(\cos \beta, g) \left[ f(\mathbf u')f(\mathbf u_*')-f(\mathbf u)f(\mathbf u_*)\right] d\mathbf \Omega d\mathbf u_*,
\end{equation}
The solution algorithm is implemented in Kinetic.jl \cite{xiao2020kinetic}, and works together with DifferentialEquations.jl \cite{rackauckas2017differentialequations} from where we are able to generate time-series data with desirable orders of accuracy along evolution trajectories.

\section{\label{sec:scheme}Solution algorithm}

\subsection{Update algorithm}

We consider a numerical algorithm within the finite volume framework.
The notation of cell-averaged particle distribution function in a control volume is adopted,
\begin{equation}
    f(t^n,\mathbf x_i,\mathbf u_j)=f_{i,j}^n=\frac{1}{\Omega_{i}(\mathbf x)\Omega_{j}(\mathbf u)} \int_{\Omega_{i}} \int_{\Omega_{j}} f(t^n,\mathbf x,\mathbf u) d\mathbf xd\mathbf u,
\end{equation}
where $\Omega_{i}$ and $\Omega_{j}$ are the cell area in the discrete physical and velocity space.
The update of distribution function can be formulated as
\begin{equation}
    f_{i,j}^{n+1}=f_{i,j}^n+\frac{1}{\Omega_{i}}\int_{t^n}^{t^{n+1}} \sum_{r=1}^{n_f} F_r \Delta S_r dt+ \int_{t^n}^{t^{n+1}} Q(f_{i,j}) dt,
    \label{eqn:update}
\end{equation}
where $F_r$ is the time-dependent flux function of distribution function at cell interface, $\Delta  S_r$ is the interface area and $n_r$ is the number of interfaces per cell.

\subsection{Interface flux}

For the numerical flux evaluation, we first reconstruct the particle distribution function around the cell interface, e.g. around $\mathbf x_{i+1/2}$,
\begin{equation}
\begin{aligned}
    &f_{i+1/2,j}^L = f_{i,j} , \\
    &f_{i+1/2,j}^R = f_{i+1,j} ,
\end{aligned}
\end{equation}
with first-order accuracy and
\begin{equation}
\begin{aligned}
    &f_{i+1/2,j}^L = f_{i,j} + \nabla_\mathbf x f_{i,j} \cdot (\mathbf x_{i+1/2}-\mathbf x_i), \\
    &f_{i+1/2,j}^R = f_{i+1,j} + \nabla_\mathbf x f_{i+1,j} \cdot (\mathbf x_{i+1/2}-\mathbf x_{i+1}),
\end{aligned}
\end{equation}
with second-order accuracy,
where $\nabla_\mathbf x f$ is the reconstructed gradient with limiters.

The interface distribution function is defined in an upwind way, i.e.,
\begin{equation}
    f_{i+1/2,j}=f_{i+1/2,j}^L H\left[\mathbf u_j\right] + f_{i+1/2,j}^R (1-H\left[\mathbf u_j\right]),
    \label{eqn:interface distribution}
\end{equation}
where $H[x]$ is the heaviside step function.
The corresponding numerical flux of particle distribution function can be evaluated via
\begin{equation}
    F_{i+1/2,j}= f_{i+1/2,j} \mathbf n_{i+1/2} \cdot \mathbf u_j,
\end{equation}
where $\mathbf n_{i+1/2}$ is the unit normal vector of cell interface and $\mathbf u_j$ denotes discrete velocity at $j$-th quadrature point.

\subsection{Collision term}

The collision term inside each cell is
\begin{equation}
    Q(f_{i,j})=\nu_i (\mathcal M_{i,j} - f_{i,j}) + \mathrm{NN}_\theta(\mathcal M_{i,j} - f_{i,j}).
\end{equation}
The collision frequency is defined as,
\begin{equation}
    \nu = p / \mu,
\end{equation}
where $p$ is pressure., and $\mu$ is viscosity coefficient.
It follows the variational hard-sphere (VHS) model's rule,
\begin{equation}
    \mu = \mu_\mathrm{ref} \left(\frac{T}{T_\mathrm{ref}}\right)^\omega,
\end{equation}
where $\mu_\mathrm{ref}$ and $T_\mathrm{ref}$ are the viscosity and temperature in the reference state, and $\omega$ is the viscosity index.

Once the interface fluxes are defined, the solution algorithm in Eq.(\ref{eqn:update}) becomes
\begin{equation}
    f_{i,j}^{n+1}=f_{i,j}^n+\frac{1}{\Omega_{i}}\int_{t^n}^{t^{n+1}} \sum_{r=1}^{n_f} F_r \Delta S_r dt+ \int_{t^n}^{t^{n+1}} \left[ \nu_i (\mathcal M_{i,j} - f_{i,j}) + \mathrm{NN}_\theta(\mathcal M_{i,j} - f_{i,j}) \right] dt.
\end{equation}
The most straightforward time-integral algorithm for the above equation is the forward Euler method.
Once the time step is much larger than mean collision time $\tau=1/\nu$, or more accurate solutions are requested, higher-order methods, e.g. the midpoint rule, Rosenbrock method \cite{shampine1982implementation}, Tsitouras's 5/4 runge-kutta method \cite{tsitouras2011modified}, can be employed.

\section{\label{sec:experiment}Numerical experiments}

In this section, we will introduce the detailed methodology for conducting numerical experiments and the solutions to validate the current model and scheme.
Both spatially uniform and non-uniform Boltzmann equations will be considered.
For convenience, dimensionless variables will be introduced in the simulations,
\begin{equation*}
\begin{aligned}
    & \tilde{\mathbf x}=\frac{\mathbf x}{L_0}, \ \tilde{\rho}=\frac{\rho}{\rho_0}, \  \tilde{T}=\frac{T}{T_0}, \ \tilde{\mathbf u}=\frac{\mathbf u}{(2RT_0)^{1/2}}, \  \tilde{\mathbf U}=\frac{\mathbf U}{(2RT_0)^{1/2}}, \\
    & \tilde{f}=\frac{f}{\rho_0 (2RT_0)^{3/2}}, \ \tilde{\mathbf T}=\frac{\mathbf T}{\rho_0 (2RT_0)}, \ \tilde{\mathbf q}=\frac{\mathbf q}{\rho_0 (2RT_0)^{3/2}},
\end{aligned}
\end{equation*}
where $R$ is the gas constant, $\mathbf T$ is stress tensor, and $\mathbf q$ is heat flux. 
The denominators with subscript zero are characteristic variables in the reference state. 
For brevity, the tilde notation for dimensionless variables will be removed henceforth. 

\subsection{Homogeneous relaxation}

First let us consider the homogeneous relaxation of particles from an initial non-equilibrium distribution, i.e.
\begin{equation*}
    f(t=0,u,v,w)= \frac{1}{2\pi^{2/3}} (\exp(-(u - 0.99)^2) + \exp(-(u + 0.99)^2)) \exp(-v^2) \exp(-w^2).
\end{equation*}
The training set is produced by the fast spectral method, which consists a series of discrete particle distribution functions from every $\Delta t=0.2$ unit time.
The detailed computational setup is shown in Table \ref{tab:relaxation}.
Notice that the viscosity coefficient in the reference state is connected with the Knudsen number,
\begin{equation*}
\mu_{0} =  \frac{5 (\alpha + 1) (\alpha + 2) \sqrt\pi}  {4 \alpha (5 - 2 \omega) (7 - 2 \omega)}   \mathrm{Kn},
\end{equation*}
where $\{ \alpha, \omega \}$ are parameters for the VHS model.
\begin{table}[htbp]
	\caption{Computational setup for training set production in homogeneous relaxation.} 
	\centering
	\begin{tabular}{llllllll} 
		\toprule 
		$t$ & $\Delta t$ &  $u$ & $v$ & $w$ & $N_u$ & $N_v$ & $N_w$  \\ 
		\hline
		$[0,2]$ & $0.2$ & $[-5,5]$ & $[-5,5]$ & $[-5,5]$ & 80 & 28 & 28 \\ 
		\hline
		Quadrature & Kn & Pr & $\mu_{\rm ref}$ & $\alpha$ & $\omega$ & Integral\\ 
		\hline
		rectangle & 1 & 2/3 & 0.554 & 1.0 & 0.5 & Tsitouras's 5/4 \\ 
		\hline
	\end{tabular} 
	\label{tab:relaxation}
\end{table}

As the initial particle distribution along $y$ and $z$ is set as equilibrium, we are mostly concerned about the evolution in $x$ direction.
We employ neural network to conduct dimension reduction, and the corresponding universal Boltzmann equation writes,
\begin{equation*}
    h_t(t,u) =\nu (\mathcal M - h) + \mathrm{NN}_\theta(\mathcal M - h).
\end{equation*}
The reduced distribution function here is defined as
\begin{equation*}
    h(t,u)=\int \int f(t,u,v,w) dv dw,
\end{equation*}
and the training data is projected into one dimension profiles in the same way.
The neural network chain consists of two hidden dense layers with $(N_u \times 16)$ neurons each, and the $\rm tanh$ plays as the activation function.
Tsitouras's 5/4 runge-kutta method \cite{tsitouras2011modified} is again used to solve the UBE and produces the data $h_\theta$ at same instants as training set, and the loss function is evaluated through mean squared error, i.e.
\begin{equation*}
    L(\theta)=\sum_{j,n} (h_\theta-h_{\mathrm{train}})^2 (t^n,u_j).
\end{equation*}

\begin{figure}[htb!]
	\centering
	\includegraphics[width=0.5\textwidth]{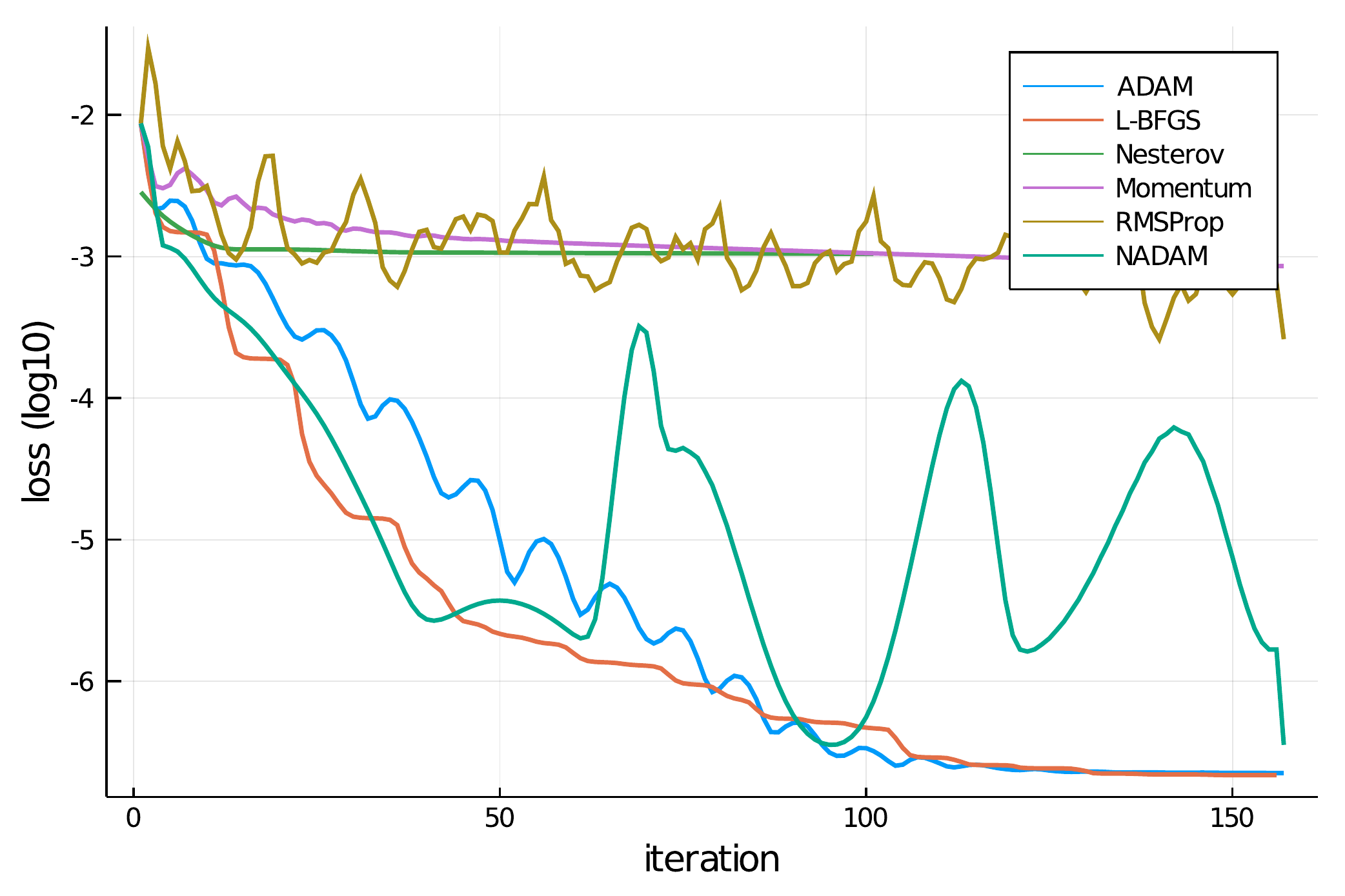}
	\caption{Loss functions of universal Boltzmann equation versus iterations under different training optimizers in the homogeneous relaxation problem.}
	\label{pic:relax loss}
\end{figure}

The variation of loss function with respect to iterations is shown in Fig.\ref{pic:relax loss}.
Different gradient-based optimizers are compared for the training of UBE.
As seen, L-BFGS enjoys the fastest convergence speed in such a supervised learning problem with relatively small amounts of data.
ADAM provides a robust gradient descent process.
Simple Momentum method, Nesterov and RMSProp, however, seems not suitable for this problem.
The hybrid Nesterov and ADAM algorithm, i.e. NADAM, presents equivalent convergence speed as L-BFGS in the beginning, but suffers from some fluctuations as the process goes.

\begin{figure}[htb!]
	\centering
	\subfigure[$t=0.2$]{
		\includegraphics[width=0.45\textwidth]{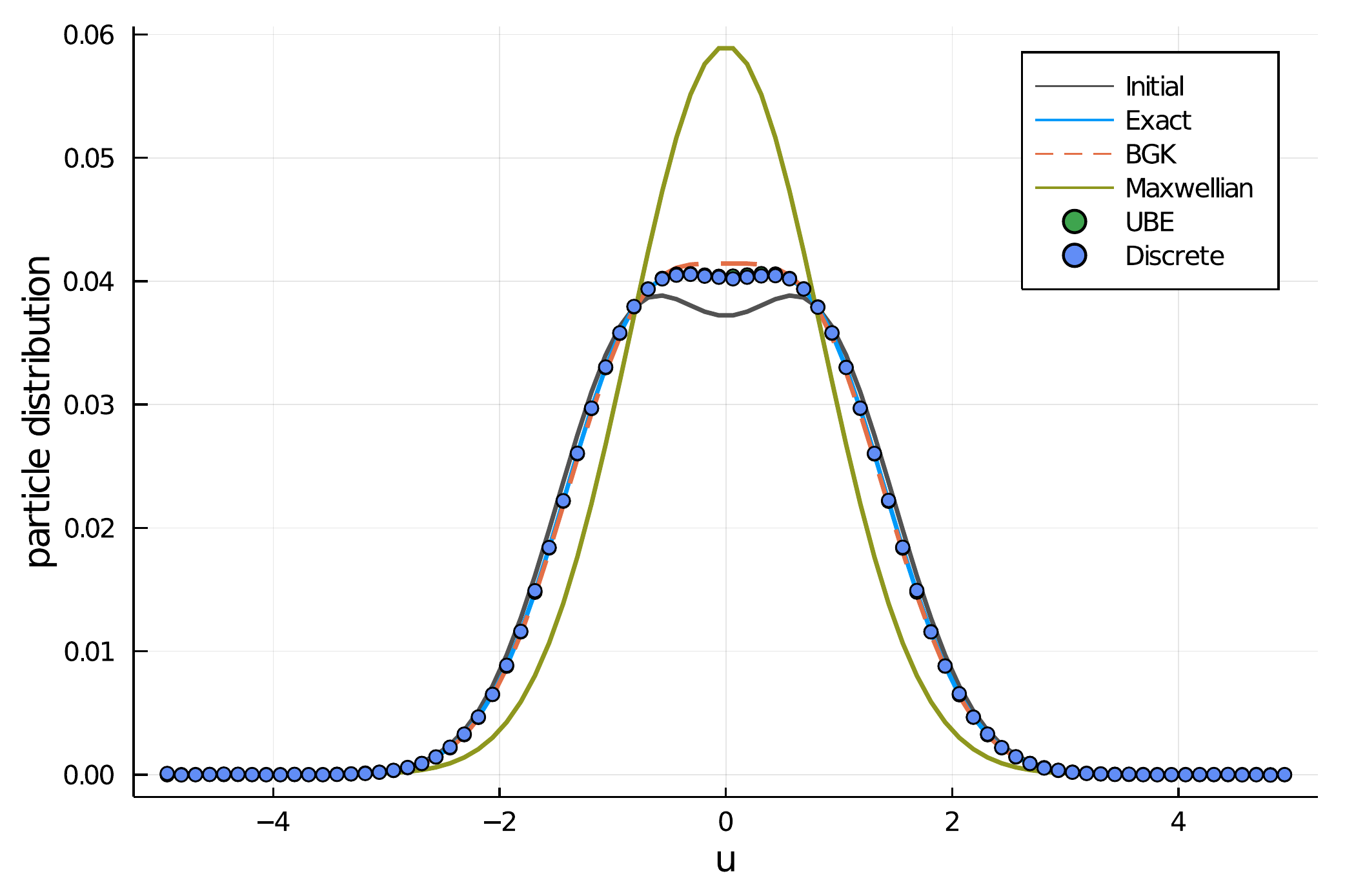}
	}
	\subfigure[$t=0.4$]{
		\includegraphics[width=0.45\textwidth]{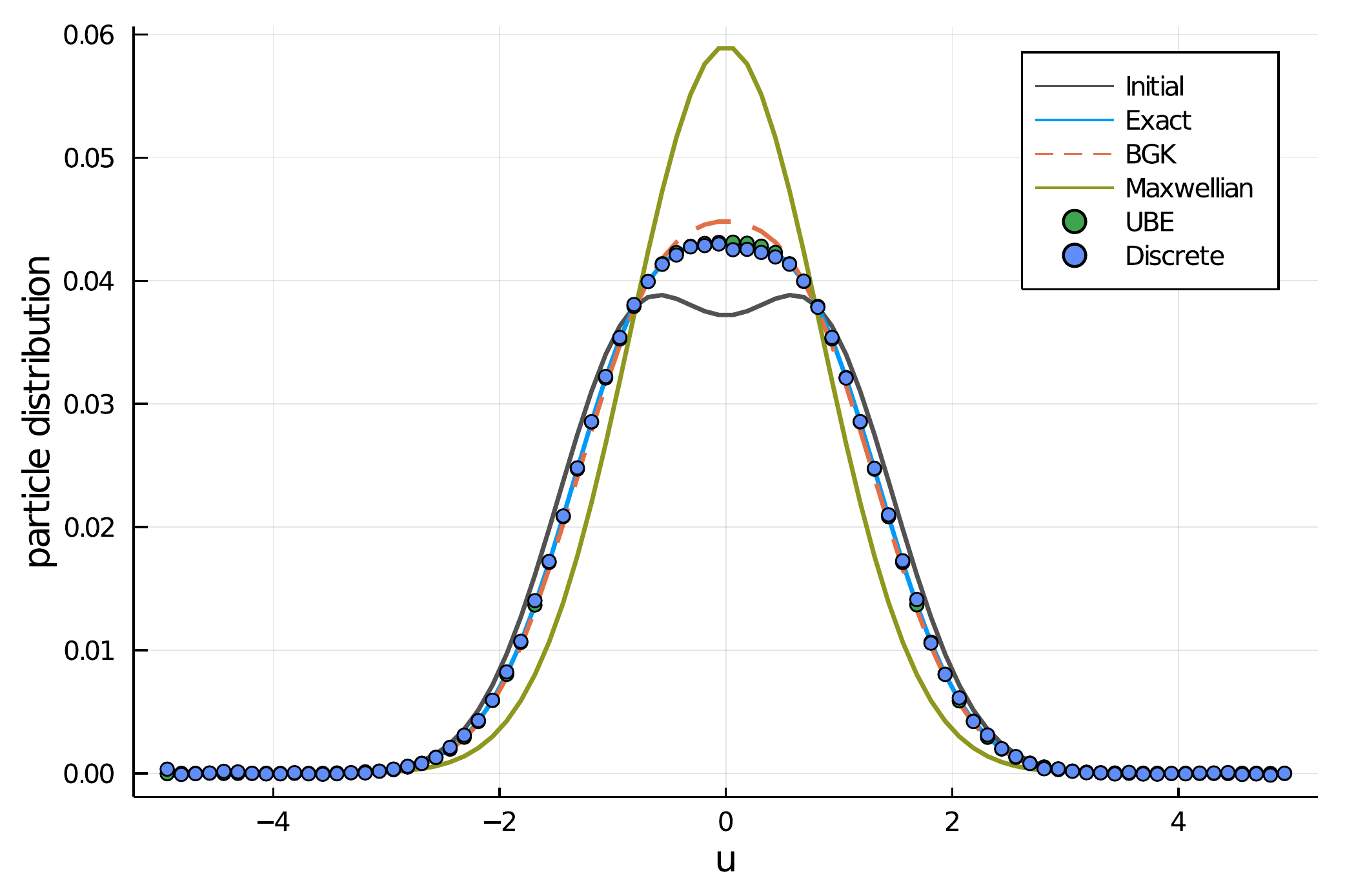}
	}
	\subfigure[$t=1.0$]{
		\includegraphics[width=0.45\textwidth]{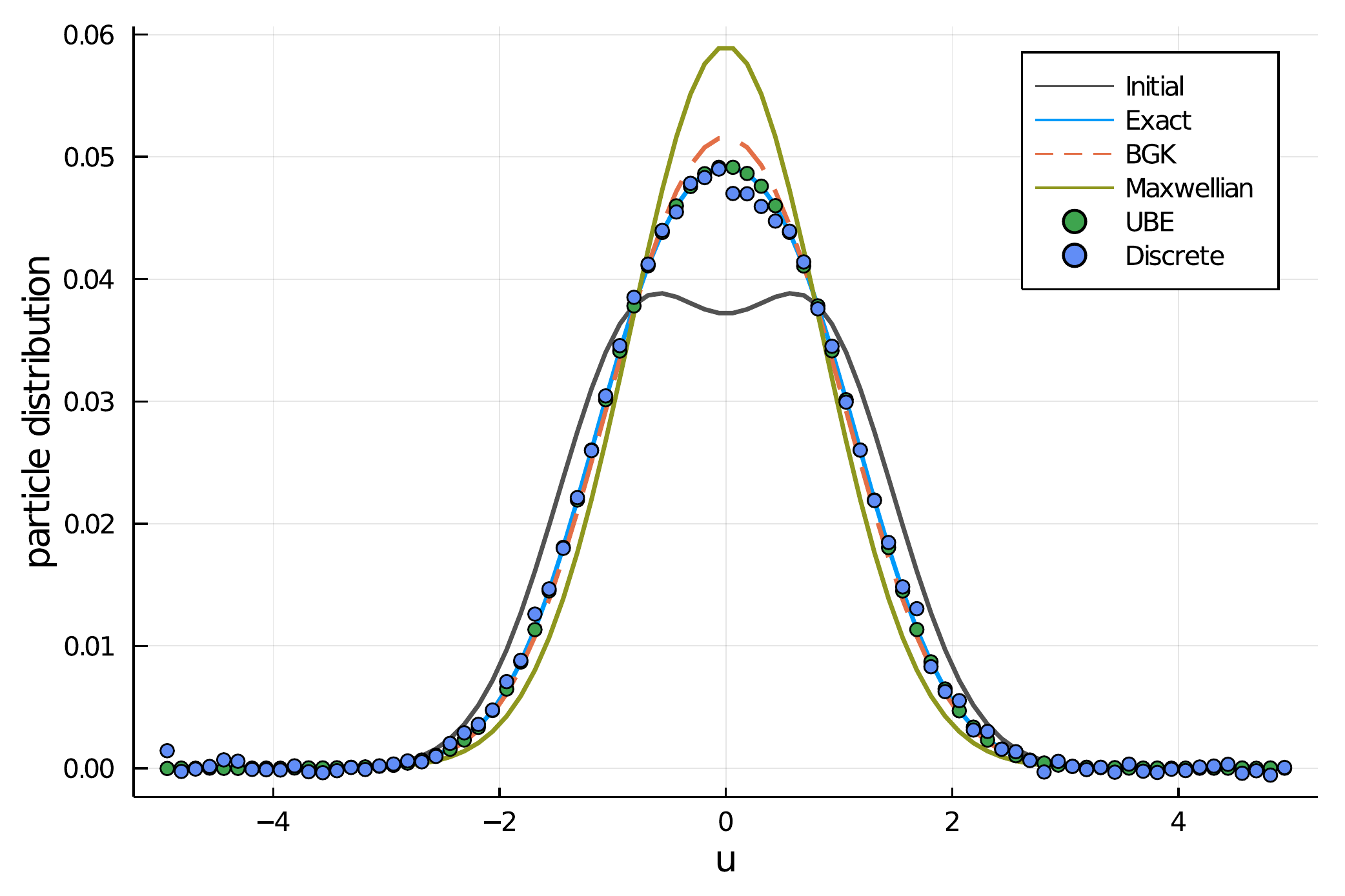}
	}
	\subfigure[$t=2.0$]{
		\includegraphics[width=0.45\textwidth]{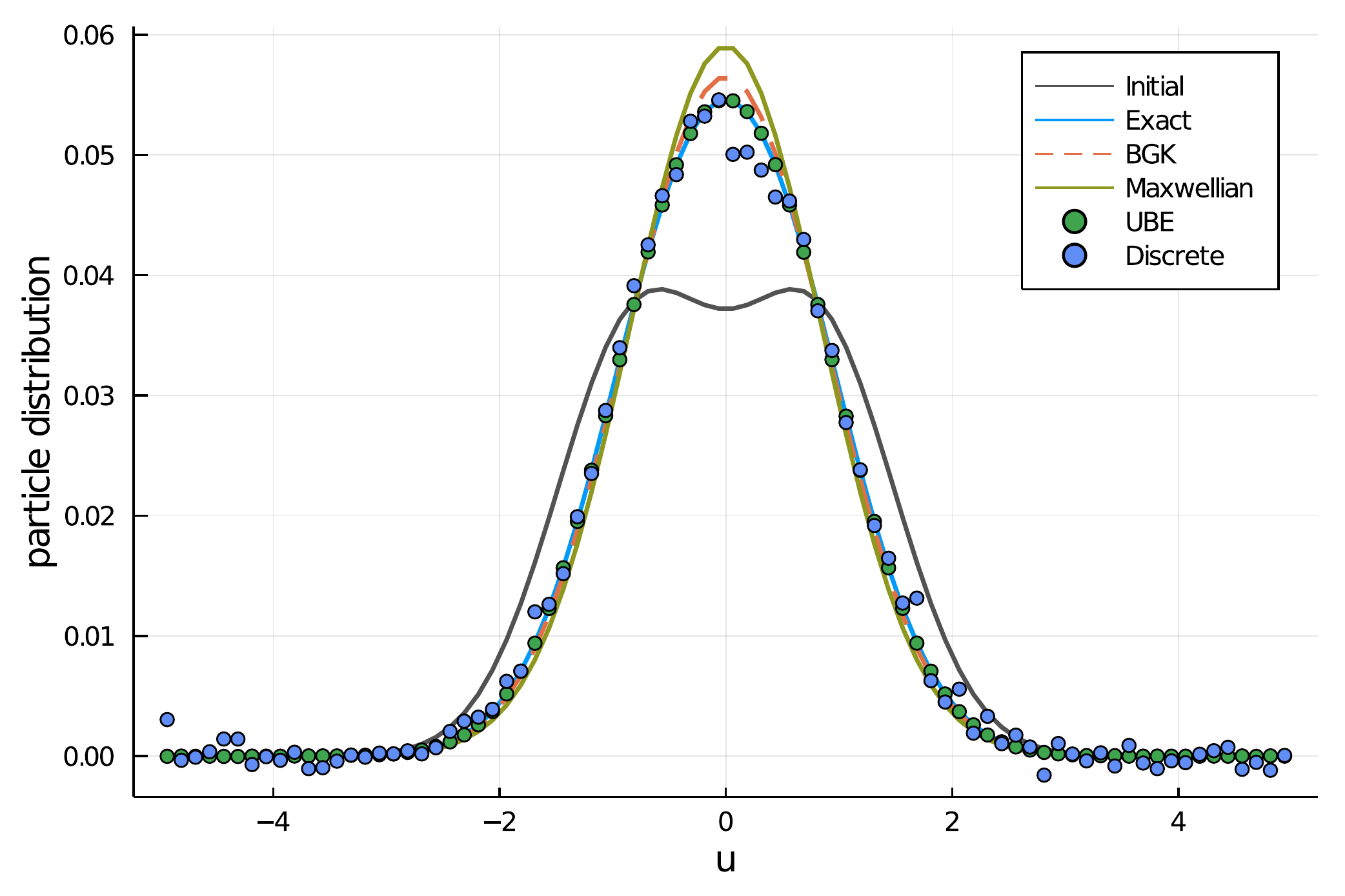}
	}
	\caption{Particle distribution functions at different time instants within the training set of the homogeneous relaxation problem.}
	\label{pic:relax train}
\end{figure}

Once the training process finishes, we will get the universal differential equation we need.
Since we are modeling continuous dynamics, instead of keeping the same approaches for producing training set, we can choose different algorithms with respect to accuracy requirement.
Hence, we use the midpoint rule to solve the obtained UBE.
Fig.\ref{pic:relax train} presents the particle distribution profile along $u$ direction at the same time instants as training set.
Different from the time-series training used in the UBE, we also plot the results with conventional discrete training strategy based on the same initial condition.
As can be seen, significant differences exist between Boltzmann and BGK solutions.
Thanks to the continuous training, the current UBE holds much better training performance based on the same datasets. 
With the mechanical part being BGK model, the UBE provides perfectly equivalent solutions as fast spectral method (FSM) of Boltzmann equation at a much lower computational cost.
Table \ref{tab:cost} shows the detailed memory usage, allocation numbers, and running time of the two methods.
As can be seen, the current method saves 97\% memory load and achieves 33 times faster computational efficiency.
\begin{table}[htbp]
	\caption{Computational cost of homogeneous relaxation.} 
	\centering
	\begin{tabular}{llllllll} 
		\toprule 
		 & MEM & $n_{allocs}$ & $t_{\rm min}$ &$t_{\rm median}$ &  $t_{\rm mean}$ & $t_{\rm max}$ & $n_{samples}$ \\ 
		\hline
		UBE &  242.25 MB & 5638 & 89.30 $ms$ & 142.43 $ms$ & 144.87 $ms$ & 221.96 $ms$ & 35\\ 
		\hline
		FSM & 7.42 GB & 194610 & 4.71 $s$ & 4.83 $s$ & 4.83 $s$ & 4.94 $s$ & 2\\ 
		\hline
	\end{tabular} 
	\label{tab:cost}
\end{table}

\begin{figure}[htb!]
	\centering
	\subfigure[$t=0.1$]{
		\includegraphics[width=0.45\textwidth]{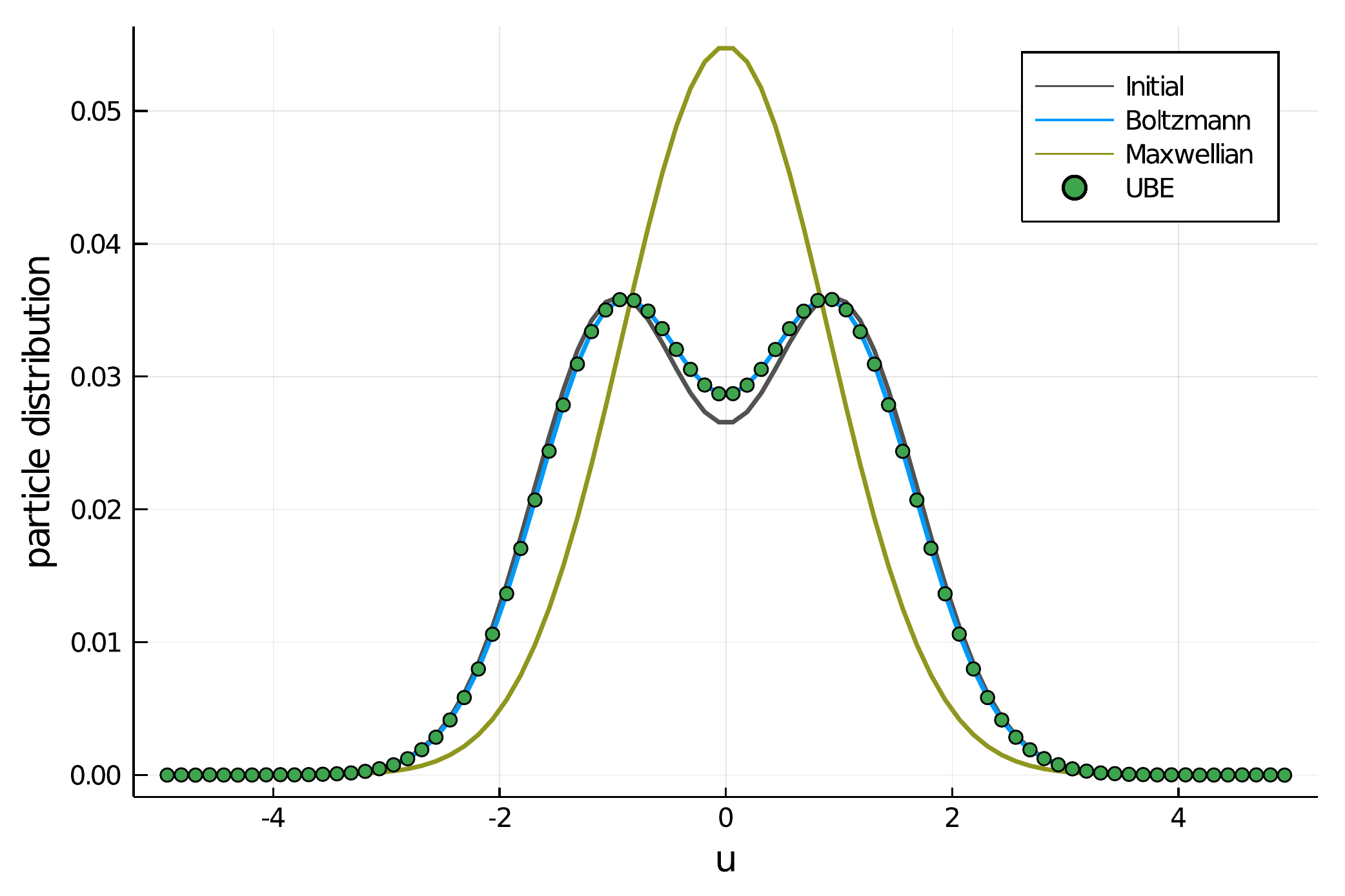}
	}
    \subfigure[$t=0.3$]{
		\includegraphics[width=0.45\textwidth]{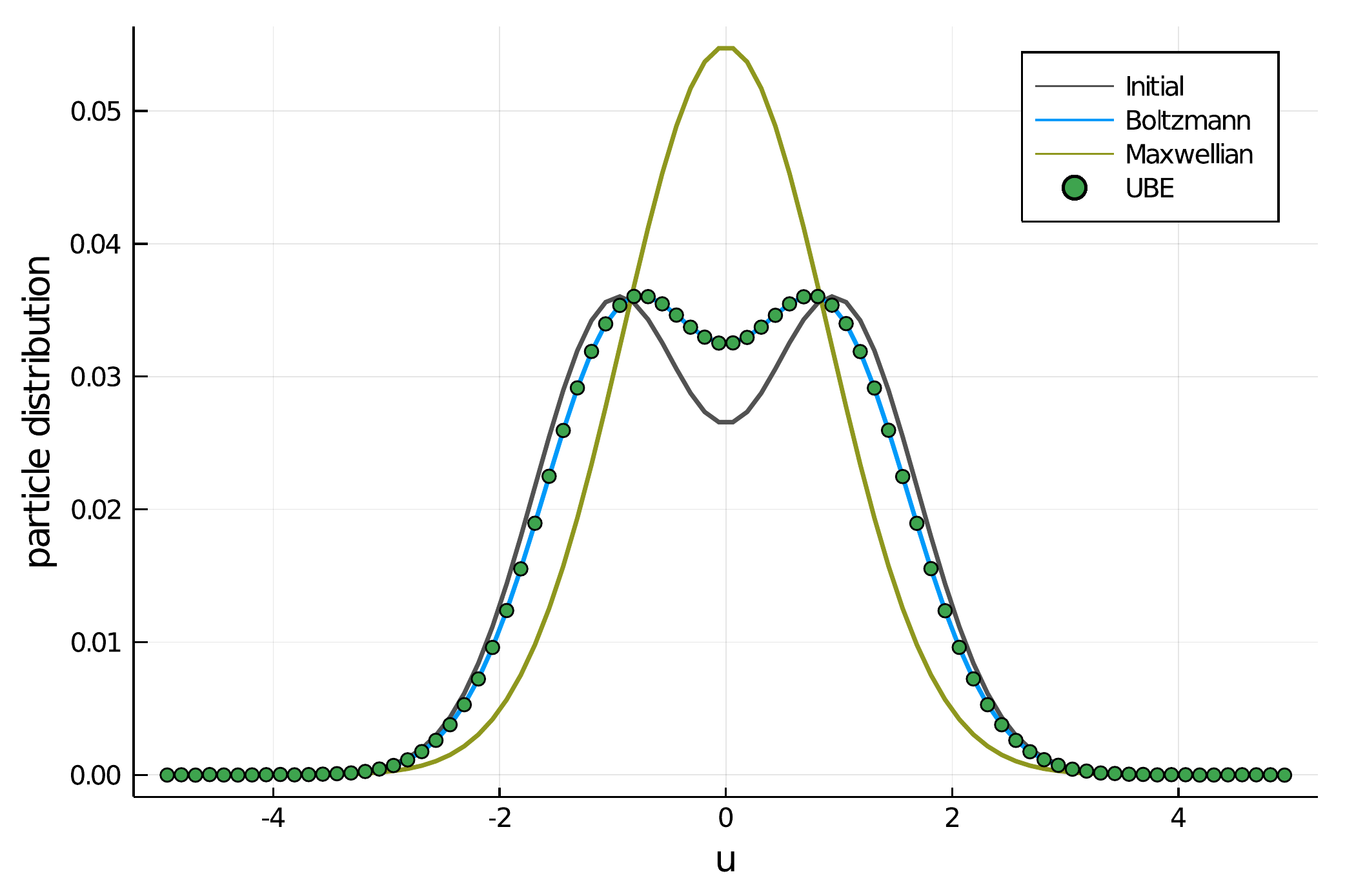}
	}
	\subfigure[$t=0.5$]{
		\includegraphics[width=0.45\textwidth]{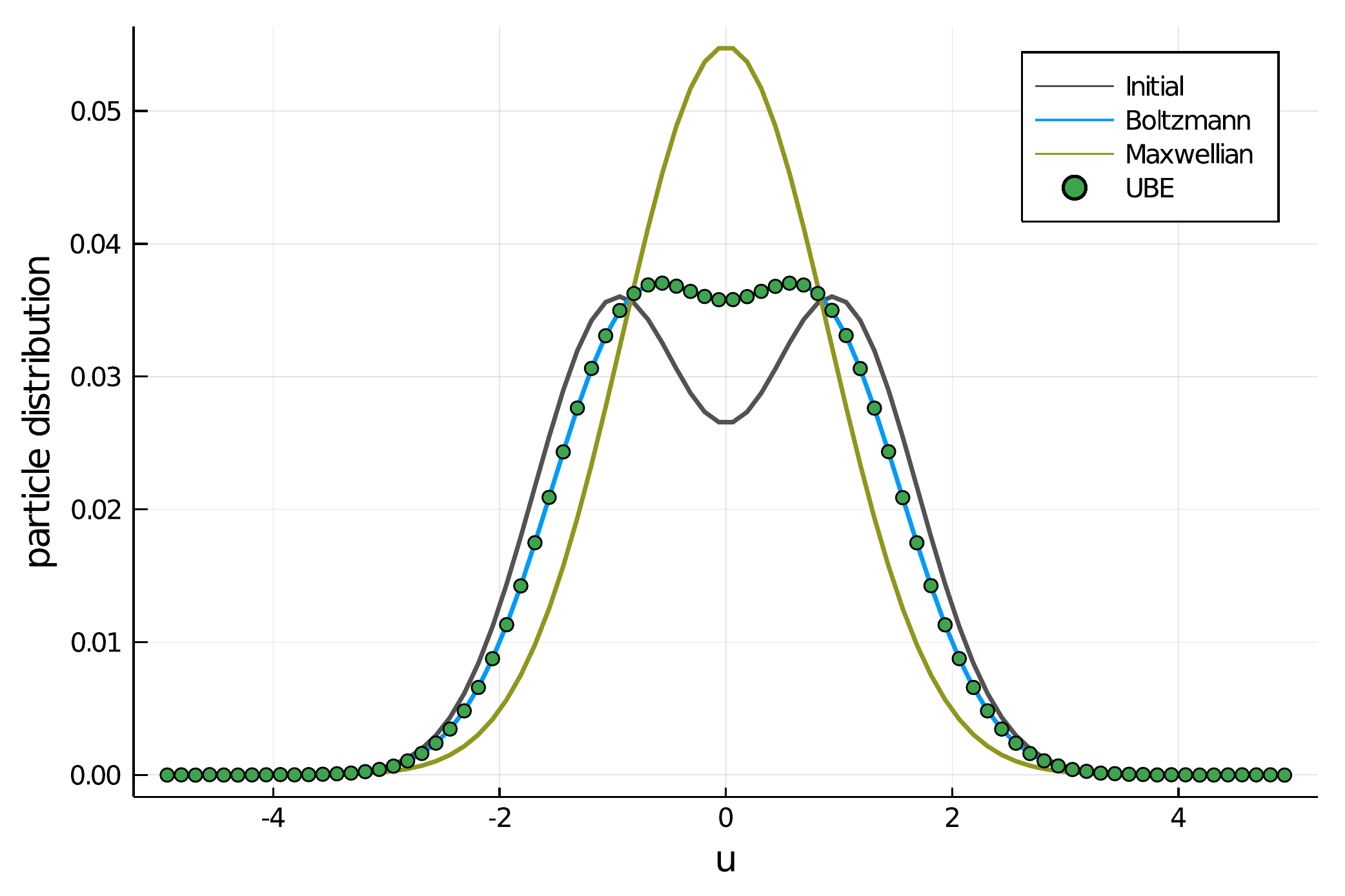}
	}
	\subfigure[$t=0.7$]{
		\includegraphics[width=0.45\textwidth]{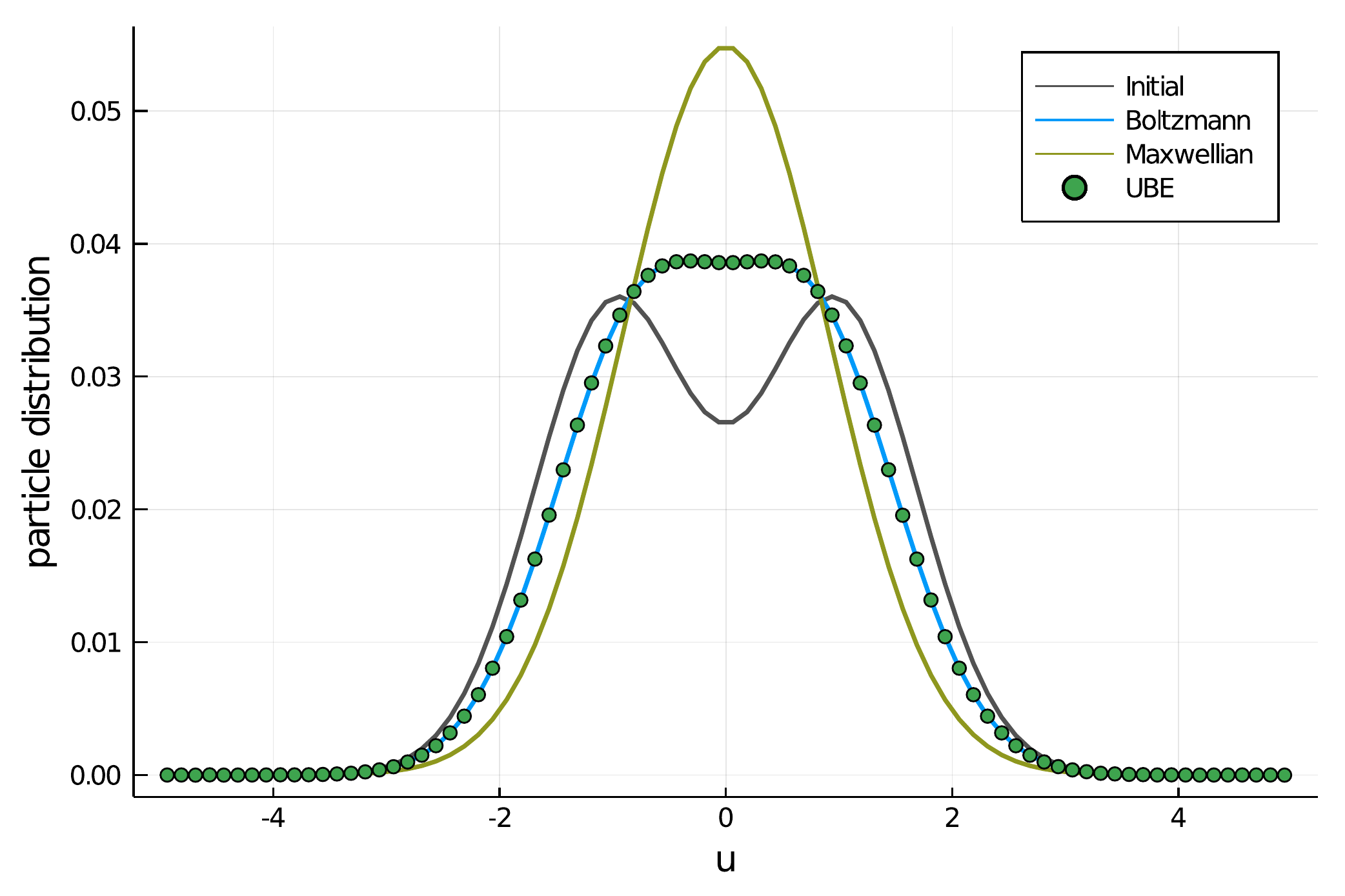}
	}
	\caption{Particle distribution functions at interpolating time instants outside the training set of the homogeneous relaxation problem.}
	\label{pic:relax inter}
\end{figure}

\begin{figure}[htb!]
	\centering
	\includegraphics[width=0.5\textwidth]{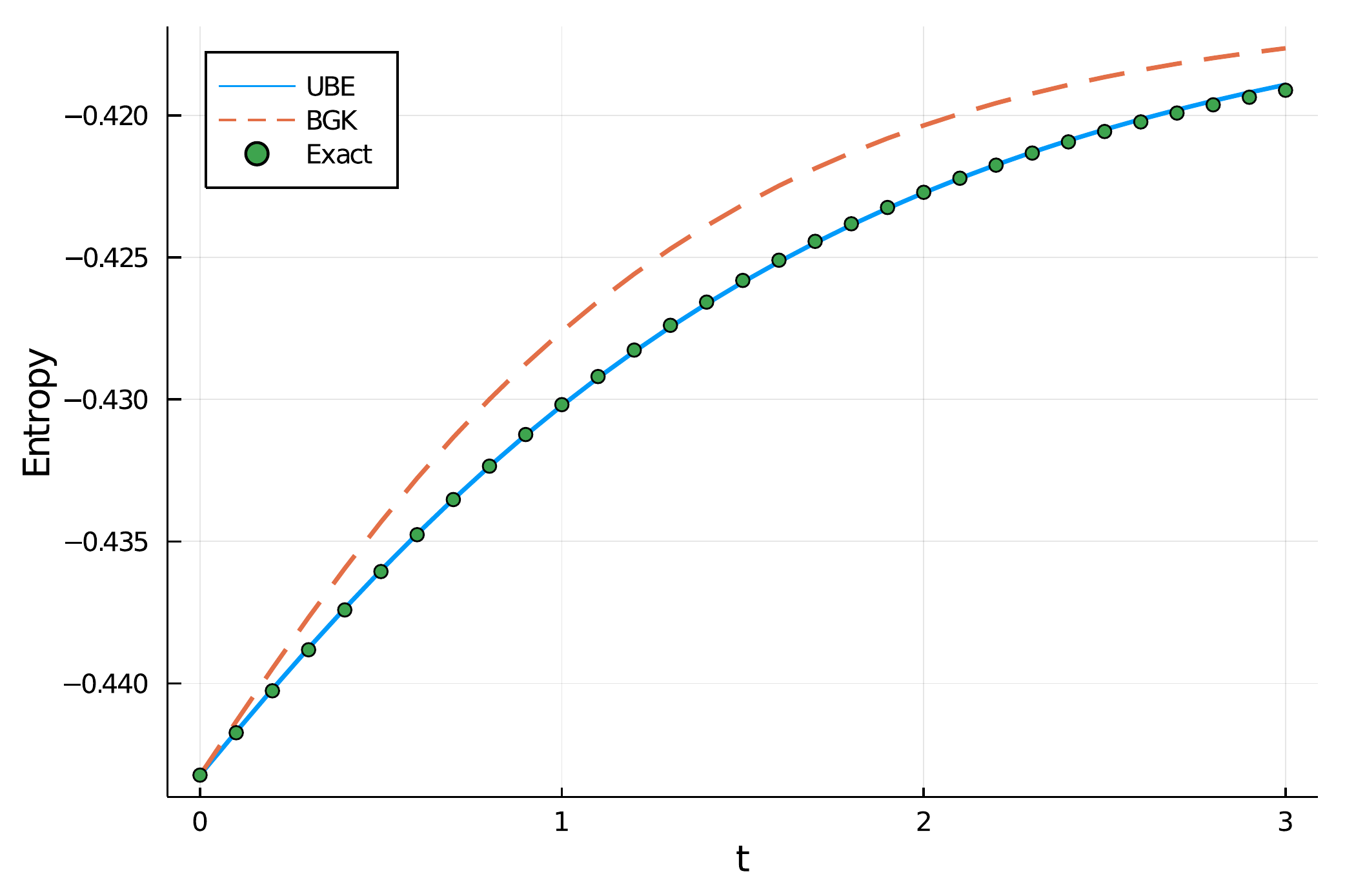}
	\caption{Time evolution of entropy in the homogeneous relaxation problem.}
	\label{pic:relax entropy}
\end{figure}

An effective neural network as function approximator should be able to conduct interpolation and extrapolation beyond the training set.
To test the performance of trained UBE, we recalculate the time-series solution within $t\in[0,9]$ and save at every $\Delta t=0.1$, 
Obviously, there exist solutions off and beyond the original training data.
Fig.\ref{pic:relax inter} presents the particle distribution profile along $u$ direction at the offset data points.
The benchmark solutions are provided by FSM with the same computational setup.
As is shown, the UBE provides equivalent solutions as FSM at interpolated and extrapolated time instants.
From Fig.\ref{pic:relax entropy}, we see that the entropy inequality is satisfied precisely by the UBE.
This case serves as a benchmark validation of the universal model and numerical scheme to provide Boltzmann solutions efficiently.

\subsection{Normal shock structure}

Then we turn to spatially inhomogeneous case.
The normal shock wave structure is an ideal case to validate theoretical modeling and numerical algorithm in case of highly disspative flow organizations and strong non-equlibrium effects.
Built on the reference frame of shock wave, the stationary upstream and downstream status can be described via the well-known Rankine-Hugoniot relation,
\begin{equation*}
\begin{aligned}
    &\frac{\rho_+}{\rho_-}=\frac{(\gamma+1)\rm{Ma}^2}{(\gamma-1)\rm{Ma}^2+2},\\
    &\frac{U_+}{U_-}=\frac{(\gamma-1)\rm{Ma}^2+2}{(\gamma+1)\rm{Ma}^2},\\
    &\frac{T_+}{T_-}=\frac{ ((\gamma-1)\rm{Ma}^2+2) (2\gamma\rm{Ma}^2-\gamma+1) }{(\gamma+1)^2 \rm{Ma}^2},
\end{aligned}
\end{equation*}
where $\gamma$ is the ratio of specific heat.
The upstream and downstream density, velocity and temperature are denoted with $\{ \rho_-, U_-, T_- \}$ and $\{ \rho_+, U_+, T_+ \}$.
The computational setup for this case is presented in Table.\ref{tab:shock}.

\begin{table}[htbp]
	\caption{Computational setup in normal shock structure.} 
	\centering
	\begin{tabular}{llllllll} 
		\toprule 
		$x$ & $N_x$ & $u$ & $N_u$ & Quadrature & Kn & Pr &   \\ 
		\hline
		$[-25,25]$ & $80$ & $[-5,5]$ & 80 & rectangle & 1 & 2/3 &  \\ 
		\hline
		$\mu_{\rm ref}$ & $\alpha$ & $\omega$ & CFL & Integral & Layer & Optimizer\\ 
		\hline
		0.554 & 1.0 & 0.5 & 0.7 & Midpoint & Dense & ADAM \\ 
		\hline
	\end{tabular} 
	\label{tab:shock}
\end{table}

\begin{figure}[htb!]
	\centering
	\includegraphics[width=0.95\textwidth]{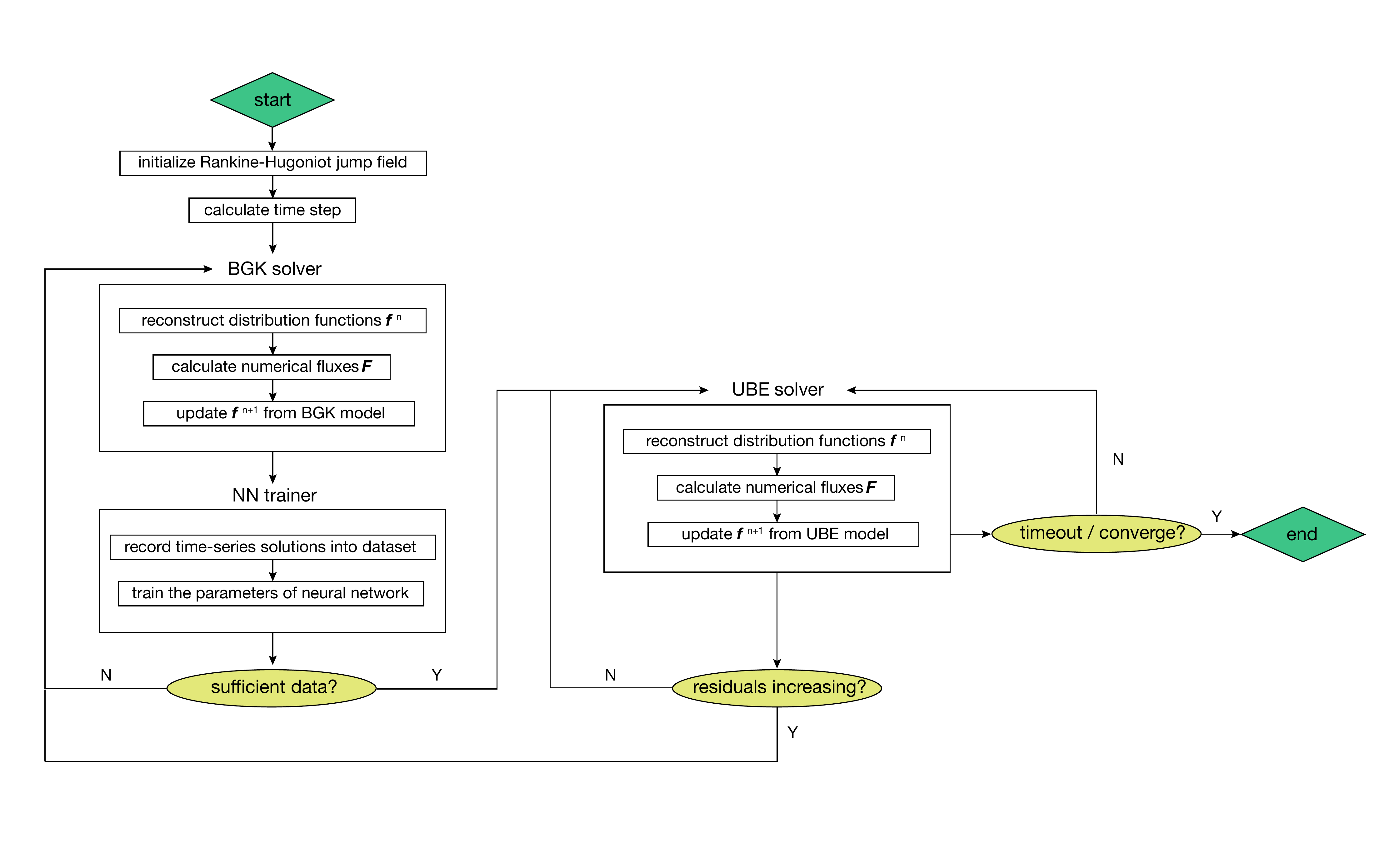}
	\caption{Training process and solution algorithm of universal Boltzmann equation in the normal shock problem.}
	\label{pic:shock train}
\end{figure}

We are only concerned about the one-dimensional profile of flow variables.
Similar as the homogeneous relaxation problem, two reduced distribution functions can be introduced to conduct dimension reduction,
\begin{equation*}
\begin{aligned}
    &h(t,x,u)=\int \int f(t,x,u,v,w) dv dw, \\
    &b(t,x,u)=\int \int (v^2+w^2) f(t,x,u,v,w) dv dw,
\end{aligned}
\end{equation*}
and the corresponding universal Boltzmann equations become,
\begin{equation*}
\begin{aligned}
    \frac{\partial h}{\partial t} + u \frac{\partial h}{\partial x} =\nu (\mathcal M_h - h) + \mathrm{NN}_\theta(\mathcal M_h - h), \\
    \frac{\partial b}{\partial t} + u \frac{\partial b}{\partial x} =\nu (\mathcal M_b - h) + \mathrm{NN}_\theta(\mathcal M_b - h).
\end{aligned}
\end{equation*}
The neural network chain consists of an input layer which accepts $h$ and $b$ with $(N_u \times 2)$ neurons, two hidden dense layers with $(N_u \times 2 \times 16)$ neurons each, $\rm tanh$ as the activation function and the output layer that is of the same shape with input.
For the current steady-state problem, BGK equation is employed first to evolve the flow field from initial jump condition, from which we extract particle distribution functions at different time instants.
As the coordinates in space results in larger data size, in this case we employ the Shakhov model to produce training set efficiently. 
The midpoint rule is employed to solve the Shakhov relaxation term and generates training data in time series \cite{xiao2017well}.
In this case five tracing solution points are recorded within each time step.
Thereafter, the same algorithm is used to solve the UBE and produces data points at the same time instants as training set.
The loss function is evaluated through mean squared error, 
\begin{equation*}
    L(\theta)=\sum_{i,j,n} (h_\theta-h_{\mathrm{train}})^2 (t^n,x_i,u_j) + \sum_{i,j,n} (b_\theta-b_{\mathrm{train}})^2 (t^n,x_i,u_j).
\end{equation*}
In the numerical simulation, the training and solving processes are handled in a coupled way.
First, the training set consists of 100 equally distributed data set (extracted every 20 time stepping and covers the entire physical domain).
After the training process, we utilize the UBE solver to continue the simulation.
However, if the residuals of flow variables keep increasing within a successive 20 time steps, we downgrade the generalization performance of the current neural network.
To overcome the overfitting of existing training data, the particle distribution functions at current time step will be extracted and added into training set to conduct parameter retraining.
The detailed training approach and solution algorithm are illustrated in Fig.\ref{pic:shock train}.

\begin{figure}[htb!]
	\centering
	\subfigure[$\rm Ma=2$]{
		\includegraphics[width=0.45\textwidth]{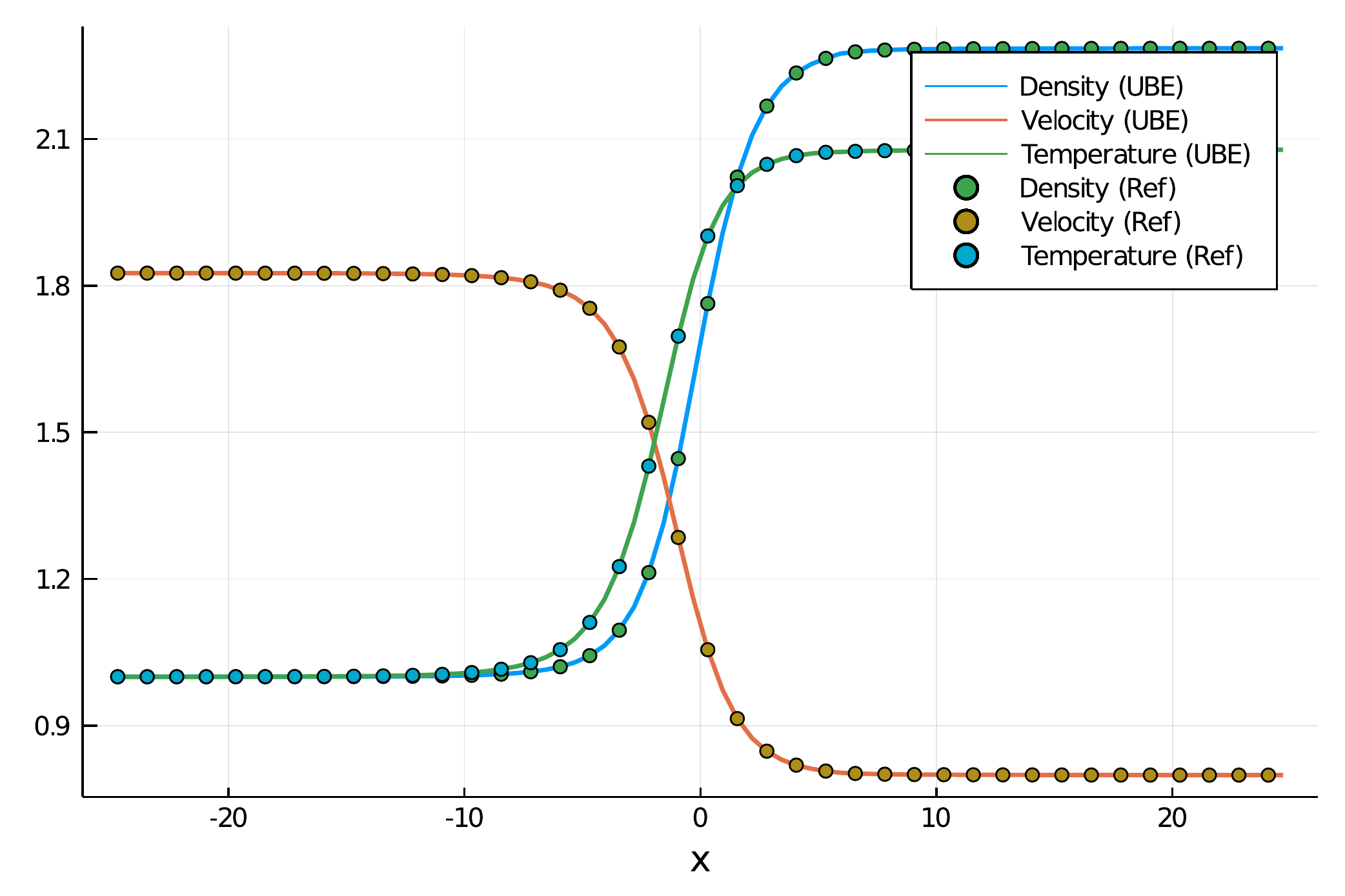}
	}
    \subfigure[$\rm Ma=3$]{
		\includegraphics[width=0.45\textwidth]{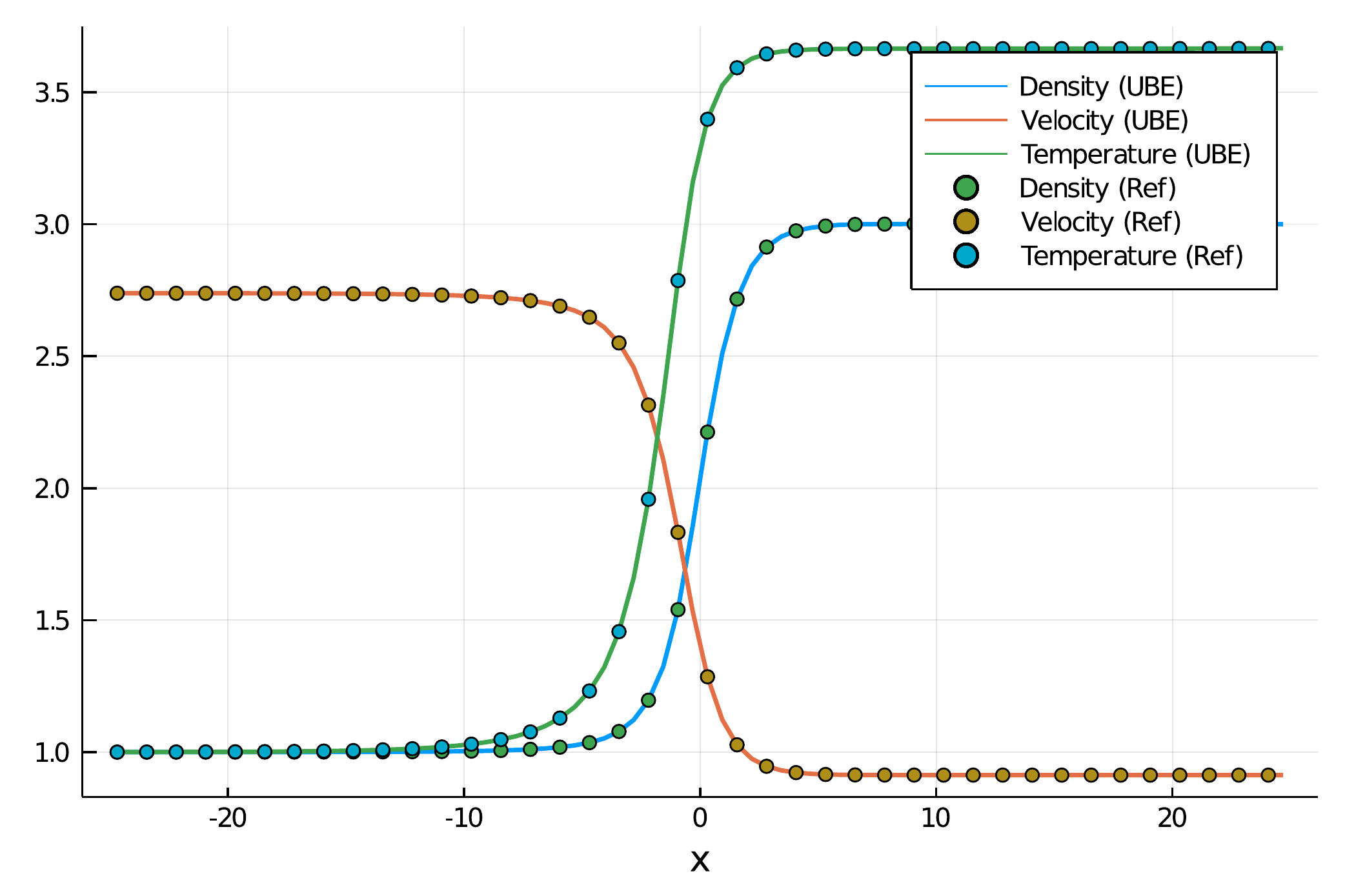}
	}
	\caption{Gas density, velocity and temperature profiles at different Mach numbers in the normal shock wave problem.}
	\label{pic:shock flow}
\end{figure}

\begin{figure}[htb!]
	\centering
	\subfigure[$\rm Ma=2$]{
		\includegraphics[width=0.45\textwidth]{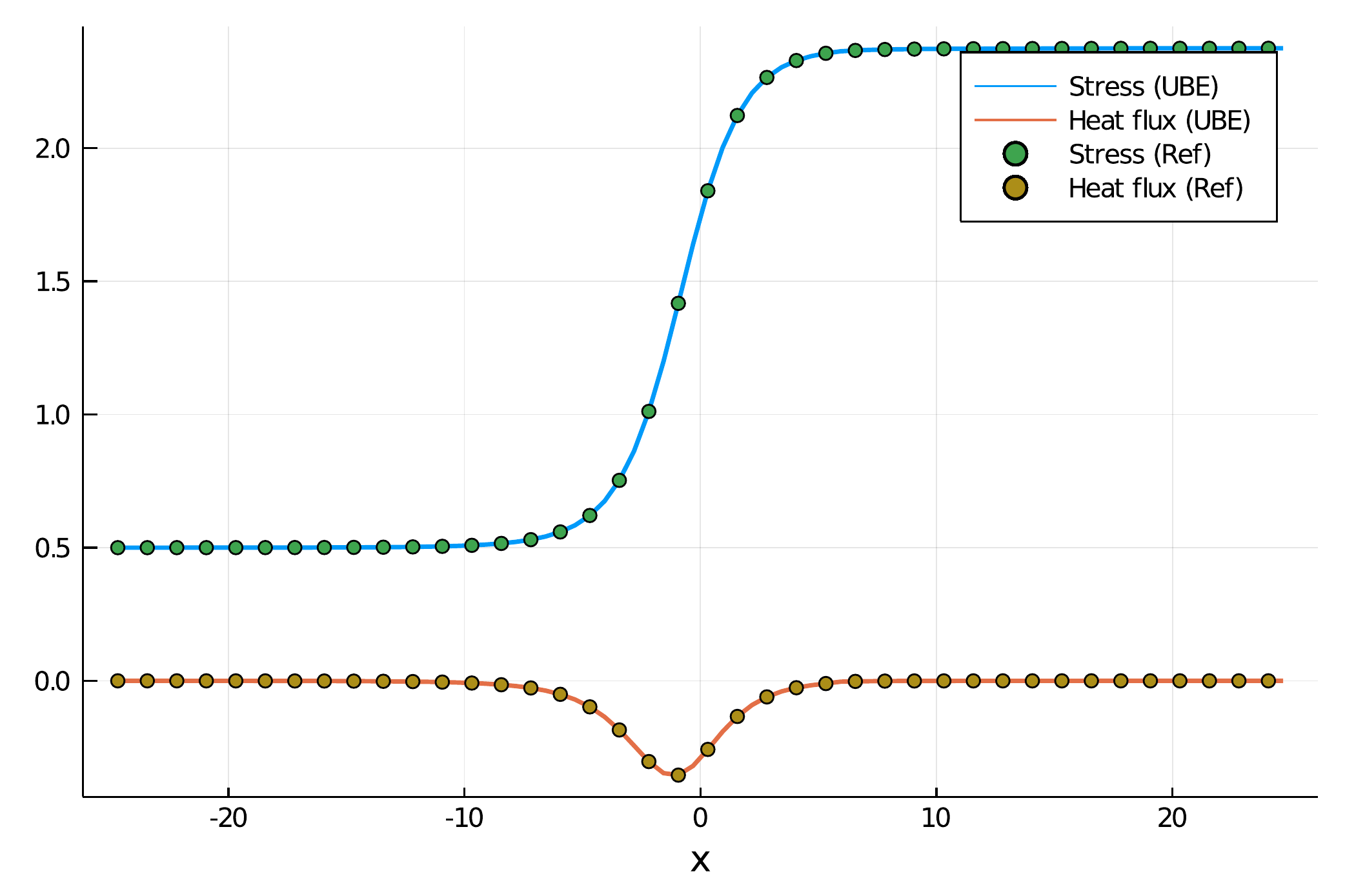}
	}
    \subfigure[$\rm Ma=3$]{
		\includegraphics[width=0.45\textwidth]{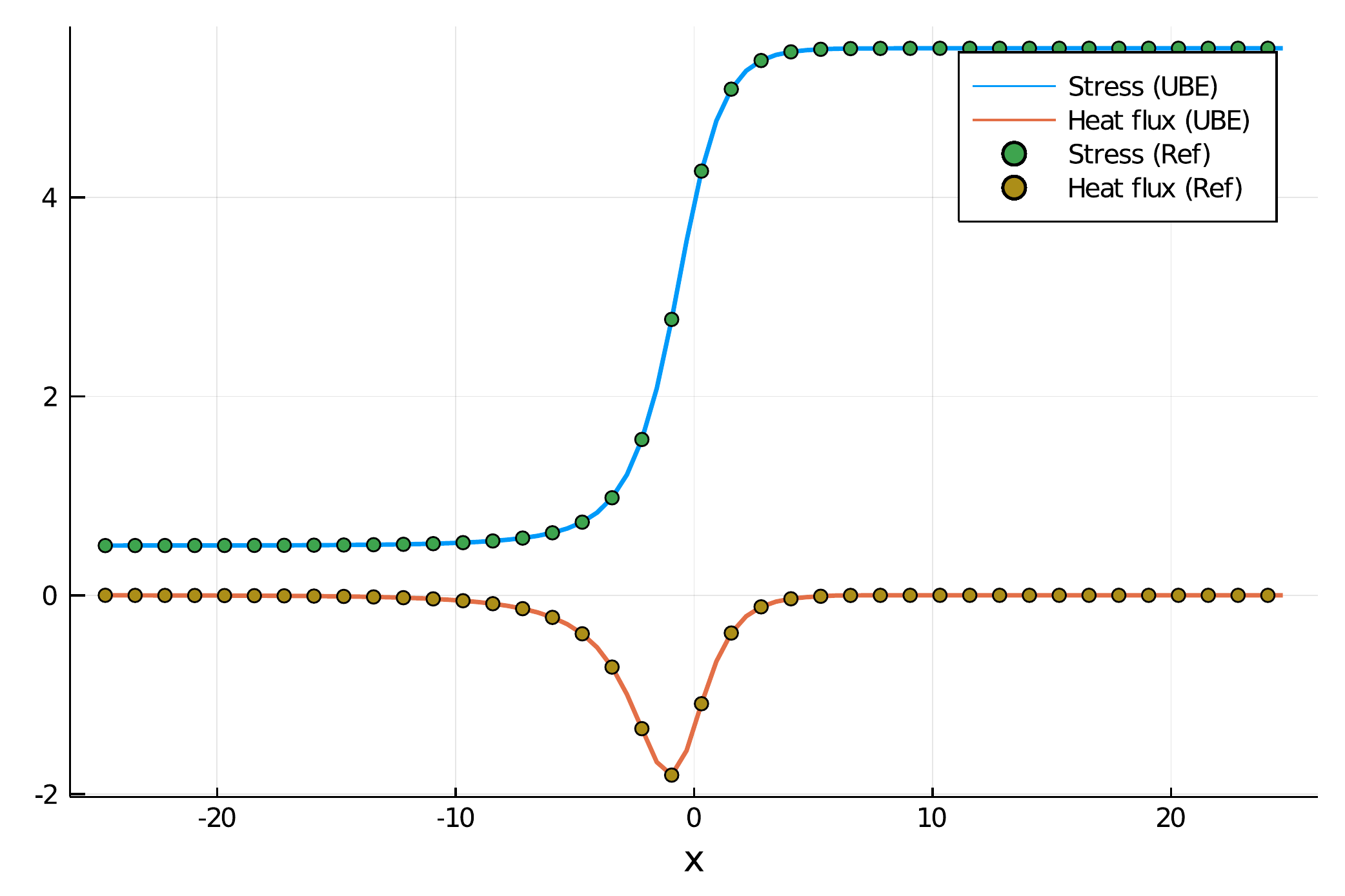}
	}
	\caption{Stress and heat flux at different Mach numbers in the normal shock wave problem.}
	\label{pic:shock therm}
\end{figure}

\begin{figure}[htb!]
	\centering
	\subfigure[Particle distribution (UBE)]{
		\includegraphics[width=0.45\textwidth]{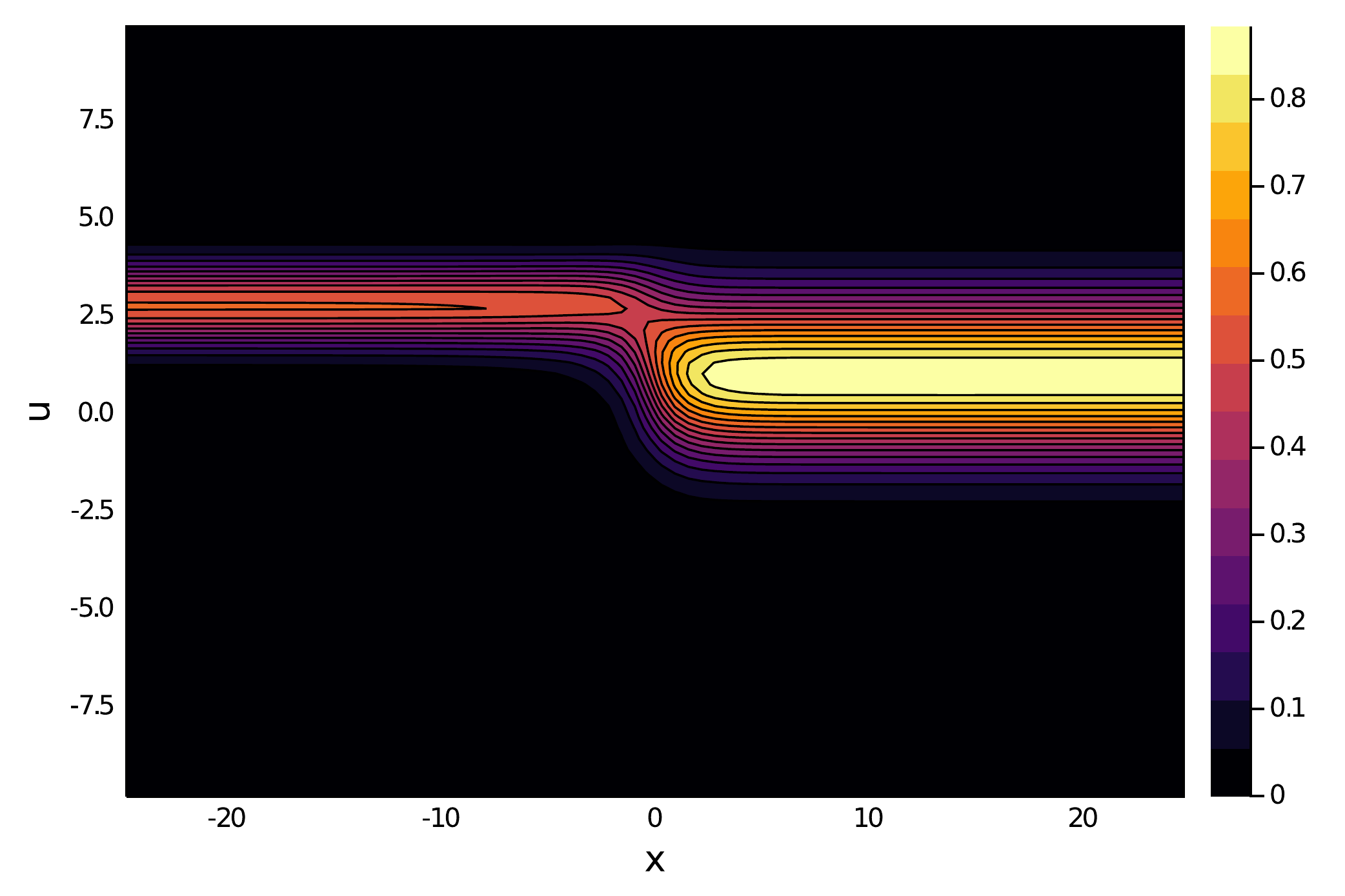}
	}
	\subfigure[Particle distribution (BGK)]{
		\includegraphics[width=0.45\textwidth]{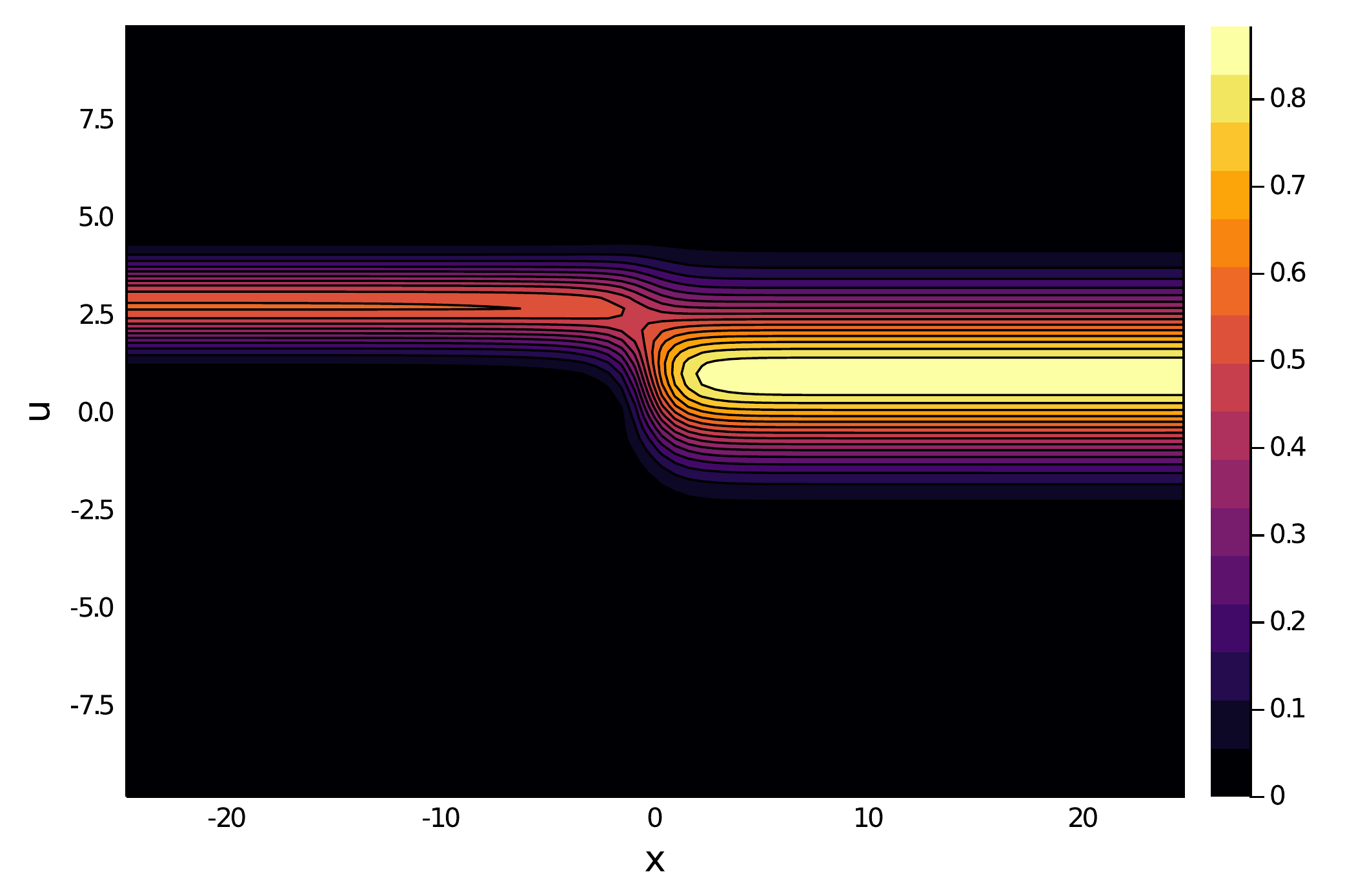}
	}
    \subfigure[Collision term (UBE)]{
		\includegraphics[width=0.45\textwidth]{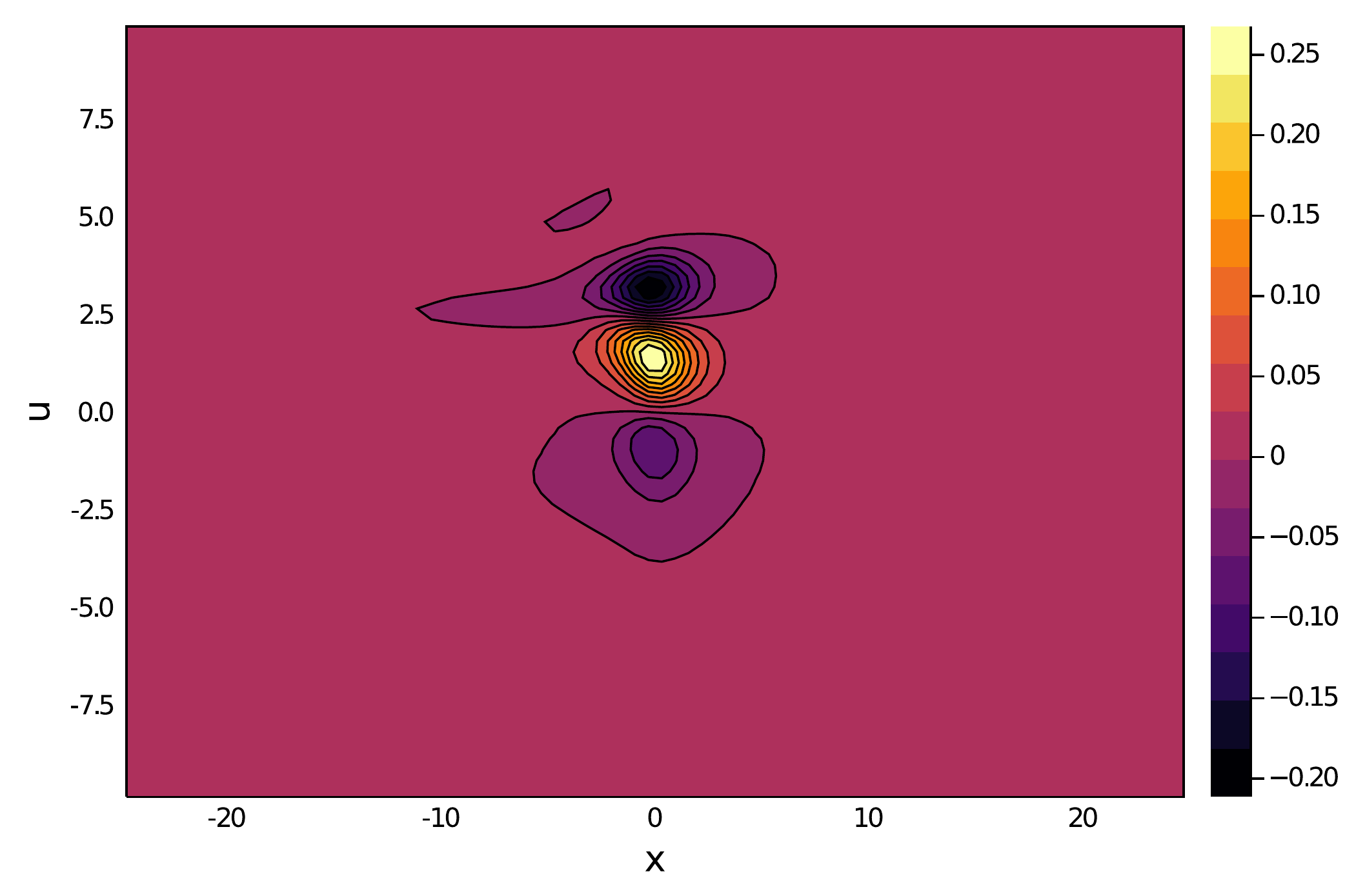}
	}
	\subfigure[Collision term (BGK)]{
		\includegraphics[width=0.45\textwidth]{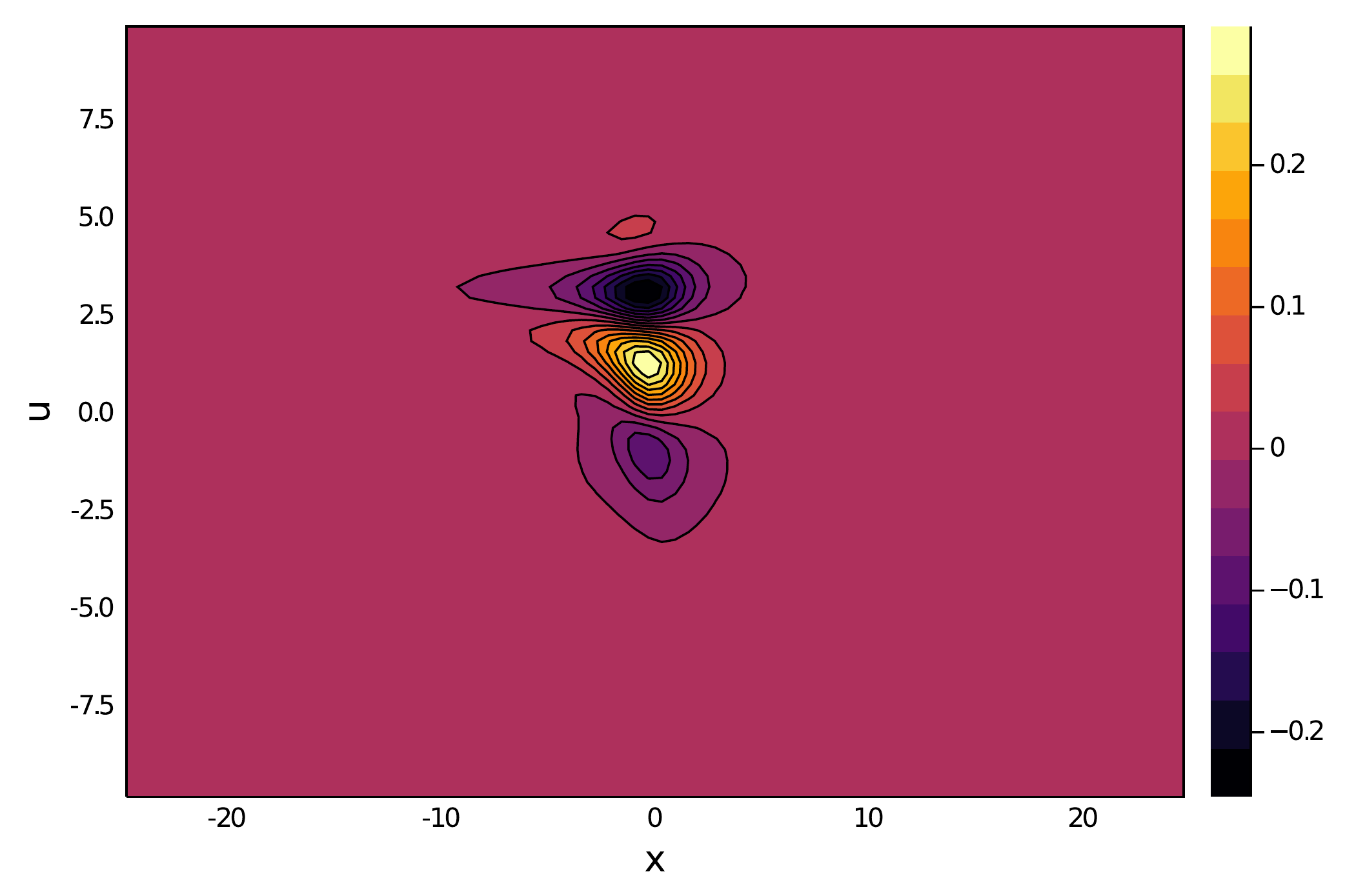}
	}
	\caption{Contours of reduced distribution functions and collision terms at $\rm Ma=3$ in the normal shock structure problem.}
	\label{pic:shock pdf}
\end{figure}

Fig.\ref{pic:shock flow} presents the profiles of gas density, velocity and temperature along $x$ direction, and Fig.\ref{pic:shock therm} provides the distributions of stress $P_{xx}$ and heat flux $q$.
As is shown, the current UBE provides equivalent solutions as reference Shakhov results.
Fig.\ref{pic:shock pdf} presents the contours of reduced particle distribution functions $h$ and collision term $\nu(\mathcal M_h-h)$ in the convergent state.
In spite of the similar patterns of particle distributions, obvious difference between UBE and BGK solution can be observed from the distribution of collision terms over the phase space $\{x,u\}$.
Table \ref{tab:shock cost} lists the detailed computational cost for evaluating collision term inside each cell.
Due to the data transmission through neurons, the UBE costs few more resources than soving Shakhov directly, but still much more efficient than the fast spectral method (FSM) for Boltzmann integral from higher dimensions in phase space.
Also, with the direct matrix manipulation possessed in neural network, fewer allocations, e.g. the collision frequency, are needed to evaluate collision terms.

\begin{table}[htbp]
	\caption{Computational cost in the normal shock structure problem.} 
	\centering
	\begin{tabular}{llllllll} 
		\toprule 
		 & MEM & $n_{allocs}$ & $t_{\rm min}$ &$t_{\rm median}$ &  $t_{\rm mean}$ & $t_{\rm max}$ & $n_{samples}$ \\ 
		\hline
		UBE &  551.20 KB & 21 & 76.10 $\mu s$ & 87.85 $\mu s$ & 126.24 $\mu s$ & 14.25 $ms$ & 10000\\ 
		\hline
		Shakhov & 4.64 KB & 30 & 3.57 $\mu s$ & 5.01 $\mu s$ & 5.29 $\mu s$ & 2.14 $ms$ & 10000\\ 
		\hline
		FSM & 63.73 MB & 1769 & 23.96 $m s$ & 27.80 $m s$ & 28.19 $m s$ & 45.60 $ms$ & 178\\
		\hline
	\end{tabular} 
	\label{tab:shock cost}
\end{table}

\section{Conclusion}

Deep learning offers another possibility for the future development of scientific modeling and simulation. 
In this paper, we hybridize mechanical and neural modelings in the context of gas dynamics and present a neural network enhanced universal Boltzmann equation (UBE).
The complicated fivefold Boltzmann integral is replaced by the neural network function approximator, forming a differentiable framework that can be trained and solved via source-to-source automatic differentiation and various differential equation solvers.
The proposed neural differential equation is a well balance of interpretability from deep physical insight and flexibility from deep learning techniques.
The asymptotic limit of the UBE in the hydrodynamic limit is preserved independent of the neural network parameters.
The solution algorithm for the UBE is provided, and numerical experiments of both spatially uniform and non-uniform cases are presented to validate the current modeling and simulation approach.
The universal Boltzmann equation method has considerable potential to be further extended to complex systems with nonelastic collisions \cite{garzo1999dense}, real-gas effects \cite{baranger2020bgk}, chemical reactions \cite{groppi1999kinetic}, uncertainty quantification \cite{xiao2020stochastic,xiao2020plasma}, etc.

\section*{Acknowledgement}

We acknowledge the help and support from Dr. Christopher Rackauckas on scientific machine learning, and the discussion with Steffen Schotthöfer.
The current research is funded by the Alexander von Humboldt Foundation.

\clearpage
\newpage

\bibliographystyle{unsrt}
\bibliography{v1}
\newpage


\begin{figure}[htb!]
	\centering
	\subfigure[$t=6.0$]{
		\includegraphics[width=0.45\textwidth]{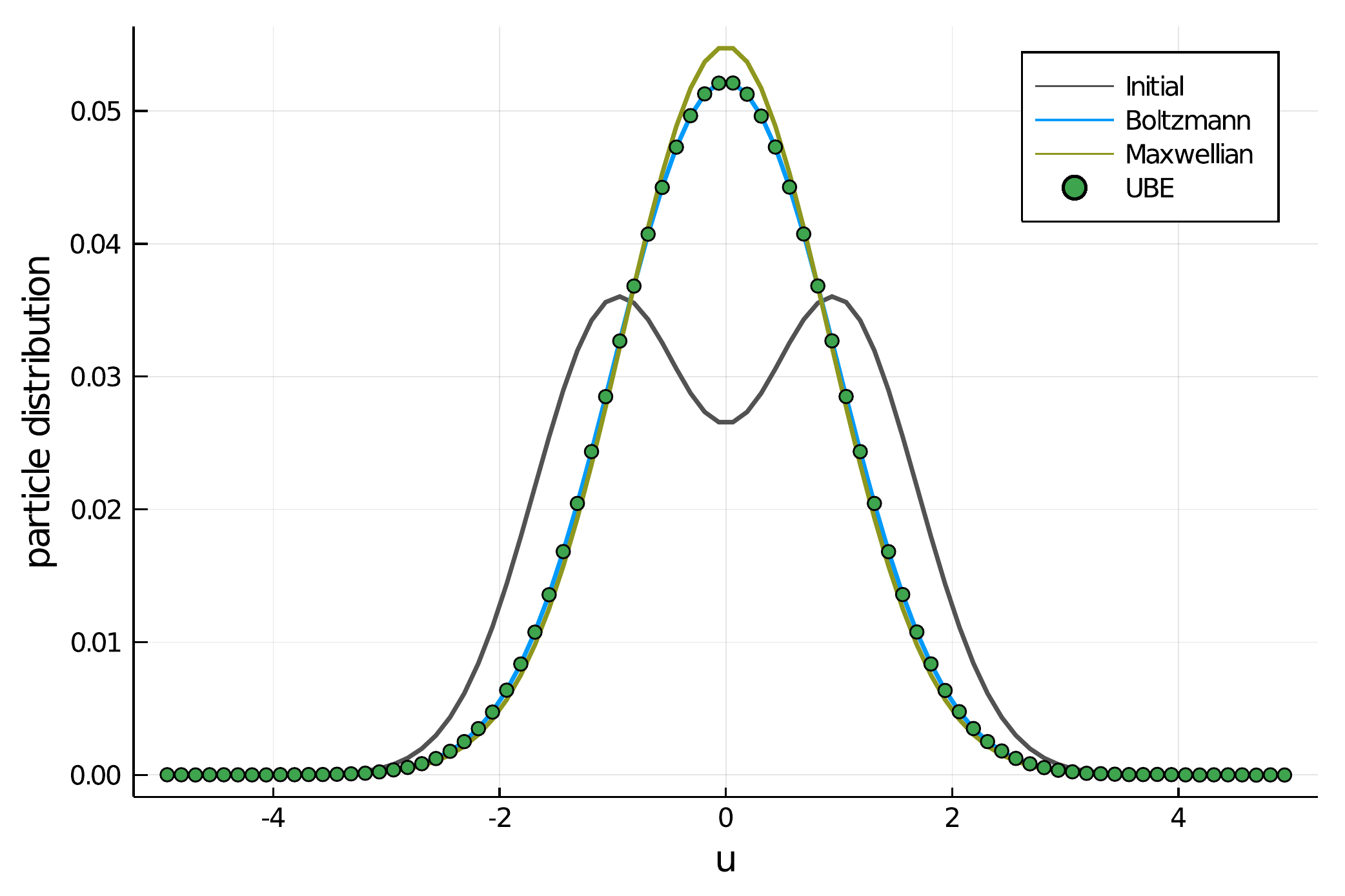}
	}
    \subfigure[$t=9.0$]{
		\includegraphics[width=0.45\textwidth]{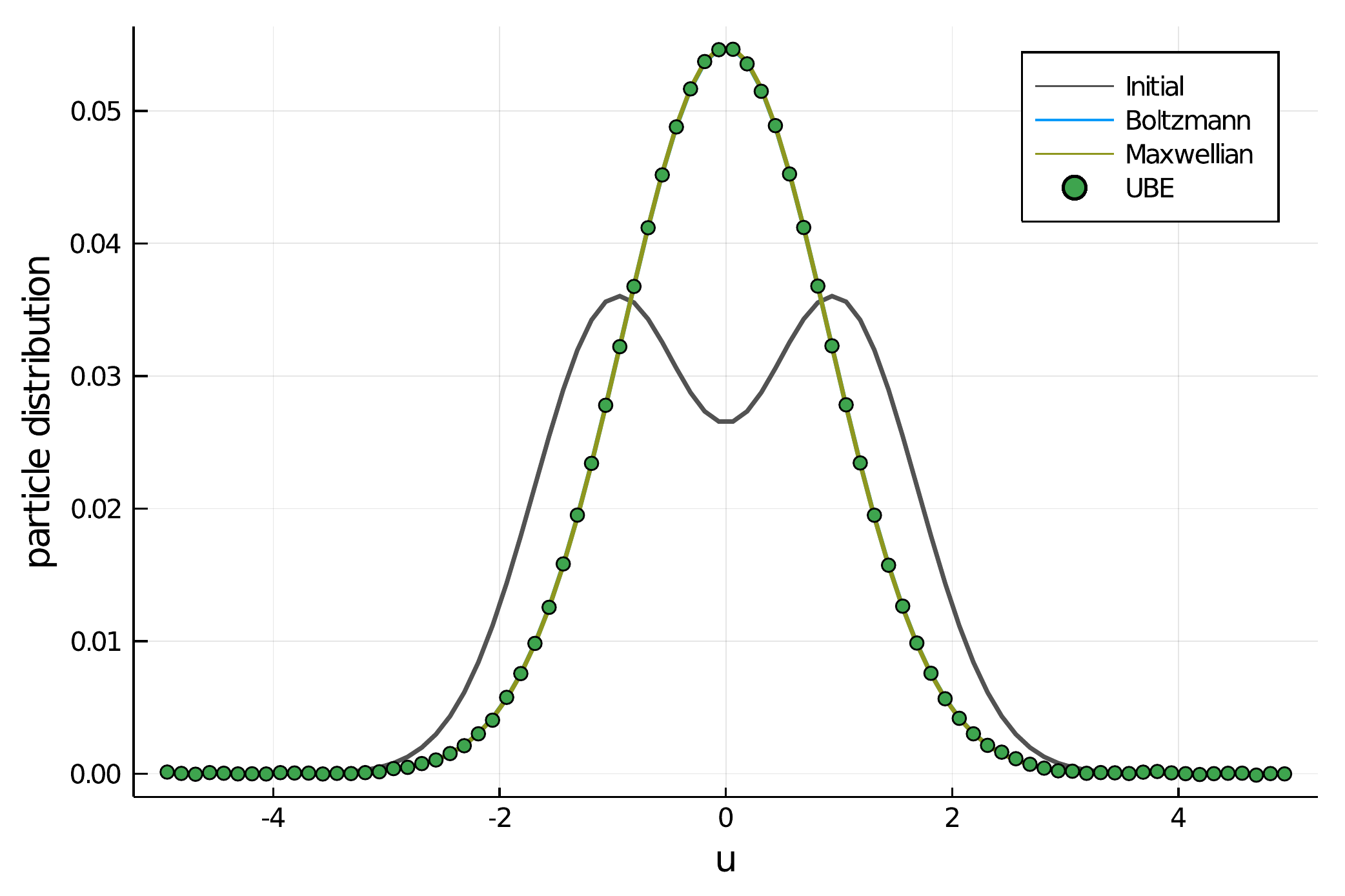}
	}
	\caption{Particle distribution functions at extrapolating time instants outside the training set of the homogeneous relaxation problem.}
	\label{pic:relax extra}
\end{figure}


\begin{figure}[htb!]
	\centering
	\subfigure[Particle distribution function]{
		\includegraphics[width=0.45\textwidth]{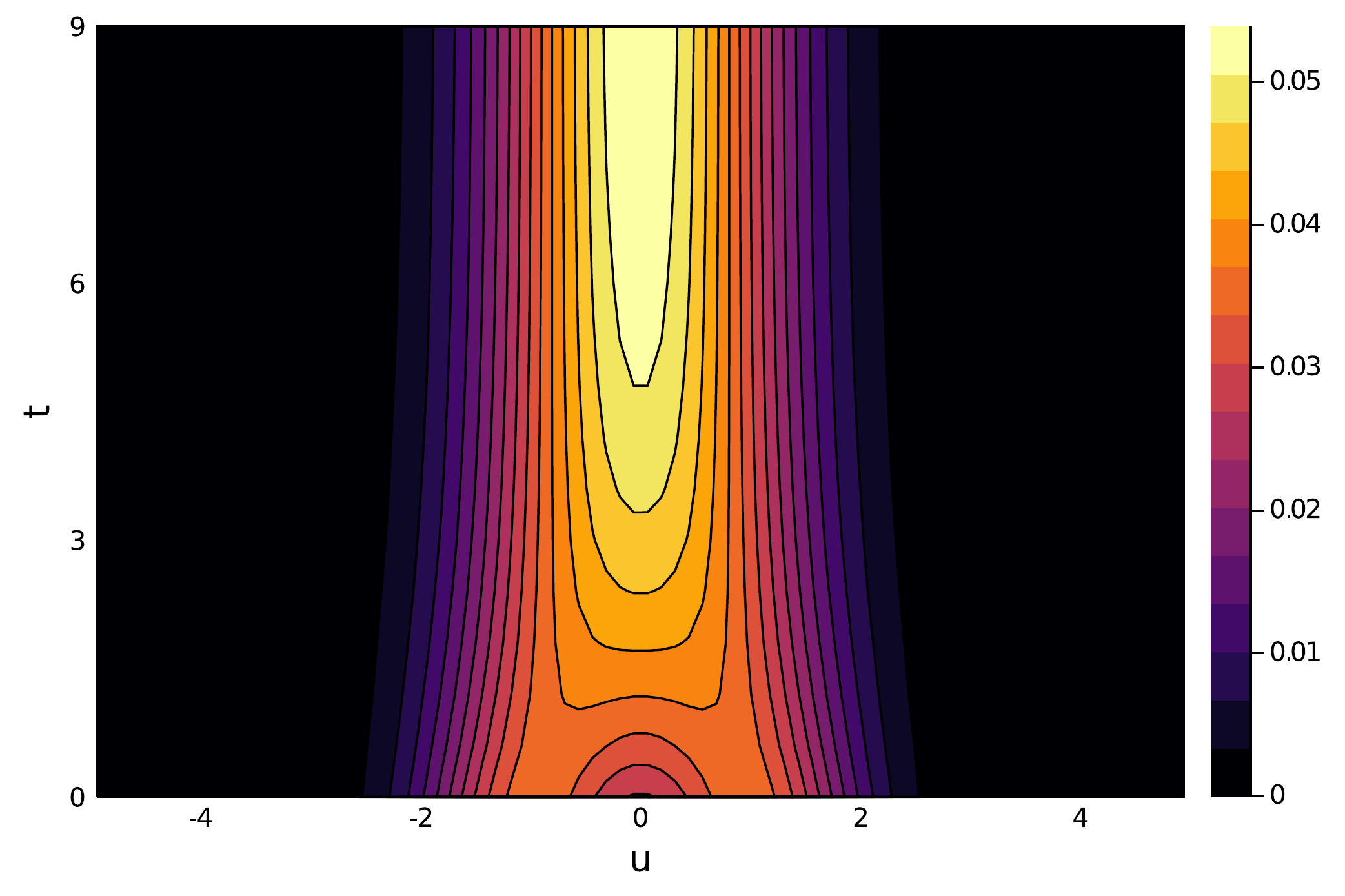}
	}
    \subfigure[Collision term]{
		\includegraphics[width=0.45\textwidth]{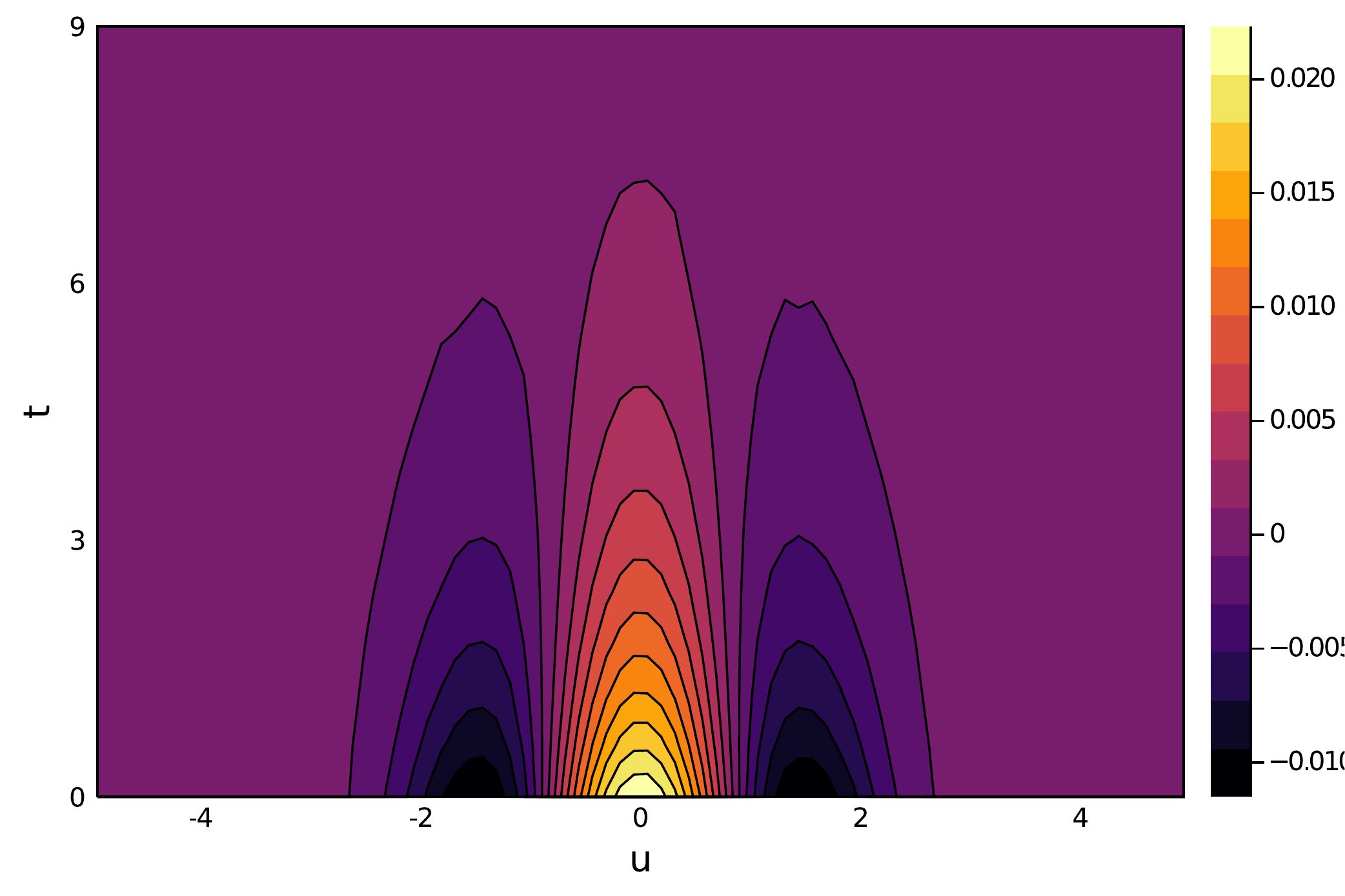}
	}
	\caption{Particle distribution functions and collision terms over phase space $\{t,u\}$ in the homogeneous relaxation problem.}
	\label{pic:relax contour}
\end{figure}

\end{document}